\documentclass[twocolumn]{aastex63}

\usepackage{gensymb}

\shortauthors{Marchand et al.}

\graphicspath{{./}{figures/}}

\begin{document}

\title{Protostellar collapse: regulation of the angular momentum and onset of an ionic precursor}

\author[0000-0002-4577-8292]{Pierre Marchand}
\affiliation{American Museum of Natural History, Department of Astrophysics, CPW at 79th, NY, 10024}
\affiliation{Department of Earth and Space Science, Osaka University, Toyonaka, Osaka 560-0043, Japan}

\author[0000-0001-8105-8113]{Kengo Tomida}
\affiliation{Department of Earth and Space Science, Osaka University, Toyonaka, Osaka 560-0043, Japan}
\affiliation{Astronomical Institute, Tohoku University, Sendai, Miyagi 980-8578, Japan}

\author[0000-0002-6907-0926]{Kei E. I. Tanaka}
\affiliation{Department of Earth and Space Science, Osaka University, Toyonaka, Osaka 560-0043, Japan}
\affiliation{ALMA Project, National Astronomical Observatory of Japan, Mitaka, Tokyo 181-8588, Japan}

\author[0000-0003-2407-1025]{Beno\^it Commer\c con}
\affiliation{CRAL, Ecole normale sup\'erieure de Lyon, UMR CNRS 5574, France}

\author[0000-0002-8342-9149]{Gilles Chabrier}
\affiliation{CRAL, Ecole normale sup\'erieure de Lyon, UMR CNRS 5574, France}
\affiliation{School of Physics, University of Exeter, Exeter, EX4 4QL, UK}

\begin{abstract}
Through the magnetic braking and the launching of protostellar outflows, magnetic fields play a major role in the regulation of angular momentum in star formation, which directly impacts the formation and evolution of protoplanetary disks and binary systems. The aim of this paper is to quantify those phenomena in the presence of non-ideal magnetohydrodynamics effects, namely the Ohmic and ambipolar diffusion. We perform three-dimensional simulations of protostellar collapses varying the mass of the prestellar dense core, the thermal support (the $\alpha$ ratio) and the dust grain size-distribution. The mass mostly influences the magnetic braking in the pseudo-disk, while the thermal support impacts the accretion rate and hence the properties of the disk. Removing the grains smaller than 0.1 $\mu$m in the Mathis, Rumpl, Nordsieck (MRN) distribution enhances the ambipolar diffusion coefficient. Similarly to previous studies, we find that this change in the distribution reduces the magnetic braking with an impact on the disk. The outflow is also significantly weakened. In either case, the magnetic braking largely dominates the outflow as a process to remove the angular momentum from the disk. Finally, we report a large ionic precursor to the outflow with velocities of several km s$^{-1}$, which may be observable.
\end{abstract}

 \section{Introduction}

Star formation takes place in molecular clouds when dense regions, the dense cores, undergo gravitational collapse. The regulation of angular momentum (AM) during this process is critical. In particular, the AM is directly linked to the question of planet formation, through the formation of a rotationally-supported protoplanetary disk, and binary star formation by disk fragmentation.
Numerous observations show that the AM is not conserved during the star formation process and that less than 1\% of the dense core's AM remains in the final star system \citep{1973asqu.book.....A,1993ApJ...406..528G,1995ARA&A..33..199B,1997ApJ...488..317O,2010A&A...512A..40M,2013EAS....62...25B}, mostly stored in the planets and companions.

Early, this phenomenon has been explained by the magnetization of the dense cores \citep{MestelSpitzer56}. Through magnetic braking, magnetic fields extract the AM from the rotating dense core and transports it into the surrounding interstellar medium (ISM). While we focus on the protostellar collapse in this study, the extraction of AM due to disk winds, photo-evaporation and disk instabilities (e.g. the Magnetorotational Instability, MRI) also occurs at later stages in the protoplanetary disk.

The first studies including magnetic fields in star-formation calculations (both analytical and numerical) were based on ideal magnetohydrodynamics (MHD). In this framework, the magnetic field is perfectly coupled to the gas and undergoes no physical dissipation. The results showed that with realistic magnetizations, dense cores being slightly unstable against collapse \citep{1999ApJ...520..706C}, the magnetic braking would remove a significant amount of AM, actually preventing the formation of large disks and strongly hindering fragmentation \citep{magneticbreakingZ,2004ApJ...616..266M,GalliShuLizano2006,PriceBate2007,HennebelleFromang2008}. This issue is known as the magnetic braking catastrophe. While the misalignement of the magnetic field with the rotation axis, or some amount of turbulence can alleviate the magnetic braking catastrophe \citep{joos,2012ApJ...747...21S,2013MNRAS.432.3320S,2013ApJ...767L..11K,2019MNRAS.489.1719W}, it became clear that a more accurate description of the magnetic field was necessary and that its decoupling to the gas could not be neglected.

The gas in dense cores is mostly composed of neutral H$_2$ and He, $\sim 70$\% and $\sim 28$\% by mass, respectively \citep{2011piim.book.....D}. The remaining $2$ \% is a mixture of heavier atoms, molecules and dust grains, either neutral or ionized. The charged species represent only a fraction of $10^{-7}$ of the particles, but the ideal MHD assumes that the collisional interactions between charged and neutral particles allow the neutrals to couple to the magnetic field whilst not being directly sensitive to the Lorentz's force. More realistically, non-ideal MHD accounts for the decoupling between the magnetic field and neutral species, as well as ions and electrons, through the ambipolar diffusion, the Hall effect and the Ohmic diffusion, respectively. In particular, ambipolar diffusion allows the neutrals to slip through the field lines, reducing the efficiency of magnetic braking. Nowadays most studies use non-ideal MHD, which seems to solve the disk and fragmentation issues of the magnetic braking catastrophe \citep{Machida_etal06,DuffinPudritz,MellonLi2009,LiKrasnopolskyShang,2015ApJ...801..117T,2016MNRAS.457.1037W,DA1,2018A&A...615A...5V}.

Beside gas, dust grains are also an important component of the ISM. These aggregates of molecules (mostly carbon and silicates) account for 1\% of the ISM mass. Because their surface is a catalyst for chemical reactions of gaseous species \citep[e.g.,][]{2015A&A...576A..49H}, and because they can hold up to several electric charges \citep{DraineSutin}, they influence the ionisation equilibrium of the gas and the nature of the ionised species. For these reasons, grains play a major role in the coupling of the gas with the magnetic field, thus being a critical factor for non-ideal MHD effects. In the diffuse ISM, the dust grain population is well described by the Mathis, Rumpl, Nordsieck (MRN) size distribution \citep{mathis}, which is commonly used in chemical calculations of non-ideal MHD resistivities \citep{2016A&A...592A..18M,2016PASA...33...41W,2019MNRAS.484.2119K,Guillet2020}. This distribution can evolve during the protostellar collapse, during which the increasing density promotes the growth of grains by accretion or coagulation \citep{1982A&A...114..245T,1991A&A...251..587R,1993ApJ...407..806C,2009A&A...502..845O,2017A&A...603A.105D}. These effects can reduce or remove smaller size grains, which is supported by observational evidences \citep{1989ApJ...345..245C,1993AJ....105.1010V}. Modeling and observing accurate dust size distributions is however a difficult challenge, and assumptions have to be made. \citet{2016MNRAS.460.2050Z} showed that the removal of smaller grains from the MRN distribution enhances the ambipolar coefficient, which effectively reduces the magnetic braking and helps the formation of large disks. The size distribution of grains is therefore an important parameter of the AM regulation during protostellar collapses.
 
Another mechanism that regulates the AM are the bipolar outflows, that are widely observed \citep{1992A&A...261..274C,1996A&A...311..858B,2002ApJ...576..222B,2017A&A...607L...6T} or produced in numerical simulations \citep{2006A&A...453..785F,2019MNRAS.486.3741H}. While the ejection of low-density fast-rotating gas makes them very efficient for the transport of specific AM, there is however no consensus on whether outflows dominate the magnetic braking to remove the normal AM from the core \citep[see for example][for two opposite conclusions]{2000ApJ...528L..41T,joos}. 

In this study, we analyse the AM regulation in various protostellar collapse simulations in the presence of ambipolar diffusion with three varying parameters, i.e., the dense core mass, the thermal support and the grain size distribution.

The paper is organised as follow. Our methods are described in Section \ref{SecMethods}, the results are presented in Section \ref{SecResults}, including analysis of the disk, outflow and AM transport. We discuss our results in Section \ref{SectDiscussion} and Section \ref{SectConclusion} is dedicated to conclusion.

\section{Simulation setup} \label{SecMethods}

\subsection{Theoretical framework}

We solve the following MHD equations

\begin{eqnarray}
  &\frac{\partial \rho}{\partial t} + \nabla \cdot \left[\rho \mathbf{u}\right]  = 0, \label{eqmass}\\
  &\frac{\partial \rho \mathbf{u}}{\partial t} + \nabla \cdot \left[\rho \mathbf{u}\mathbf{u} + \left(P+\frac{B^2}{2}\right) \mathbb{I} - \mathbf{BB} \right]  = -\rho \nabla \Phi, \label{eqmom} \\ 
  &\frac{\partial \mathbf{B}}{\partial t} - \nabla \times \left[\mathbf{u} \times \mathbf{B} - \frac{c^2}{4\pi}\eta_\Omega \mathbf{J} + \frac{c^2}{4\pi}\eta_\mathrm{AD}\frac{\left(\mathbf{J} \times \mathbf{B}\right) \times \mathbf{B}}{B^2} \right]  = 0,   \label{eqinduc} \\
  &\nabla \cdot \mathbf{B}=0.\label{nablab} 
\end{eqnarray}

$\rho$ is the gas density, $\mathbf{u}$ is its velocity, $P$ is the thermal pressure, $\mathbf{B}$ is the magnetic field, $\Phi$ is the gravitational potential, $c$ is the speed of light, $\eta_\Omega$ and $\eta_\mathrm{AD}$ are the Ohmic and ambipolar diffusion resistivities in s and $\mathbf{J}=\nabla \times \mathbf{B}$ is the electric current. We do not consider the Hall effect in this study.
The system is closed by a simple barotropic equation of state (EOS) to mimic the evolution of temperature in collapsing cores \citep{Larson1969,Masunaga2000}

\begin{equation}\label{EqEOS}
  T=T_0 \left[1+ \left(\frac{\rho}{\rho_*} \right)^{\gamma-1} \right],
\end{equation}
with $T_0=10$ K, $\rho_*=10^{-13}$ g cm$^{-3}$ and $\gamma=5/3$.

\subsection{Initial conditions}\label{Secinit}

Our setup follows the standard \citet{1979ApJ...234..289B} initial conditions of a spherical core in solid rotation along the z-axis with an azimuthal density perturbation.
We use a rotational energy to gravitational potential energy ratio of $\beta = 0.03$. The density follows a $m=2$ azimuthal mode
\begin{equation}
  \rho = \rho_0 [1 + \delta_\rho \cos (2 \phi)],
\end{equation}
with $\rho_0$ the average density, $\delta_\rho=0.1$ the amplitude of the perturbation and $\phi$ the azimuthal angle. This idealized m=2 perturbation is commonly used to emulate the low-m spiral arms produced by gravitational instability, even in the presence of turbulence inherited from the ISM at larger scales \citep{2018MNRAS.475.5618B}.
The magnetic field is initially parallel to the rotation axis. Its strength is defined using the mass-to-flux ratio over the critical value \citep{MouschoviasSpitzer1976}

\begin{equation}
  \mu=\frac{\frac{M}{\Phi_\mathrm{B}}}{\left(\frac{M}{\Phi_\mathrm{B}}\right)_\mathrm{crit}},
\end{equation}
where $M$ is the mass of the core, $\Phi_\mathrm{B}=\pi R_0^2 B_0$ is the magnetic flux, $R_0$ the initial core radius, $B_0$ the initial magnetic field strength and

\begin{equation}
  \left(\frac{M}{\Phi_\mathrm{B}}\right)_\mathrm{crit}=\frac{0.53}{3\pi}\sqrt{\frac{5}{G}}
\end{equation}
is the critical value, with $G$ the gravitational constant.
In all simulations, the initial mass-to-flux ratio of the cloud is $\mu=5$.

The varying parameters are the mass of the core ($M = 2$ or $5$ $M_\odot$), the thermal to gravitational energy ratio ($\alpha= 0.3$ or $0.4$) and the grain size distribution (see Section \ref{SecResistivities}). All cases are summarized in Table \ref{parameters}.

\begin{deluxetable*}{lcccccccc}[t]
  \caption{Simulation parameters : name of the simulation, mass of the initial sphere, thermal to gravitational energy ratio, average density, radius of the sphere, initial angular velocity, specific angular momentum, initial magnetic field and grain size distribution.}
  \label{parameters}
  \tablehead{\colhead{Name}                 & \colhead{Mass ($M_\odot$)} & \colhead{$\alpha$}   & \colhead{$\rho_0$ (g cm$^{-3}$)} & \colhead{$R_0$ (au)} & \colhead{$\Omega_0$ (rad s$^{-1}$)} & \colhead{$\ell$} (cm$^2$ s$^{-1}$) & \colhead{$B_0$ ($\mu$G)}   & \colhead{Grain size distribution}}
\startdata
  M2a3mrn       &  $2$             &   $0.3$    & $1.38 \times 10^{-18}$ & $5900$   & $1.8\times 10^{-13}$    & $5.93 \times 10^{20}$ & 66.7         & Standard MRN \\
  M2a3big       &  $2$             &   $0.3$    & $1.38 \times 10^{-18}$ & $5900$   & $1.8\times 10^{-13}$    & $5.93 \times 10^{20}$ & 66.7         & Truncated MRN\\
  M2a4mrn       &  $2$             &   $0.4$    & $5.74 \times 10^{-19}$ & $7900$   & $1.2\times 10^{-13}$    & $6.84 \times 10^{20}$ & 37.5         & Standard MRN  \\
  M2a4big       &  $2$             &   $0.4$    & $5.74 \times 10^{-19}$ & $7900$   & $1.2\times 10^{-13}$    & $6.84 \times 10^{20}$ & 37.5         & Truncated MRN  \\
  M5a3mrn       &  $5$             &   $0.3$    & $2.17 \times 10^{-19}$ & $14830$  & $7.8\times 10^{-14}$    & $1.48 \times 10^{21}$ & 26.7         & Standard MRN \\
  M5a3big       &  $5$             &   $0.3$    & $2.17 \times 10^{-19}$ & $14830$  & $7.8\times 10^{-14}$    & $1.48 \times 10^{21}$ & 26.7         & Truncated MRN \\
  M5a4mrn       &  $5$             &   $0.4$    & $9.15 \times 10^{-20}$ & $19770$  & $4.8\times 10^{-14}$    & $1.70 \times 10^{21}$ & 15.0         & Standard MRN \\
  M5a4big       &  $5$             &   $0.4$    & $9.15 \times 10^{-20}$ & $19770$  & $4.8\times 10^{-14}$    & $1.70 \times 10^{21}$ & 15.0         & Truncated MRN \\
\enddata
\end{deluxetable*}

In this paper, we refer the ``formation time of the first Larson core" $t_\mathrm{C}$ as the time at which the maximum density reaches $10^{-13}$ g cm$^{-3}$, and ``the first core" as all the gas that exceeds that density.
We define the protostellar disk in the same manner as \citet{joos}, as the gas verifying the following conditions 

\begin{itemize}
  \item[-] $\rho > 3.8 \times 10^{-15}$ g cm$^{-3}$,
  \item[-] $u_\phi > 2u_r$, with $u_\phi$ and $u_r$ the azimuthal and radial velocities,
  \item[-] $u_\phi > 2u_z$, with $u_z$ the velocity along the $z$ axis,
  \item[-] $\frac{\rho u_\phi^2}{2} > P$ to ensure that the gas is supported against gravity by its centrifugal force rather than the thermal pressure.
\end{itemize}

We also define an outflow as the gas verifying the two following conditions

\begin{itemize}
  \item[-] The velocity is higher than the escape velocity, $||\mathbf{u}|| > u_\mathrm{lib} = \sqrt{\frac{2GM_\mathrm{core}}{r}}$, where $M_\mathrm{core}$ is the mass of the first Larson core and $r$ the distance to the center of first core.
  \item[-] The radial velocity is positive $u_r > 0$.
\end{itemize}

\subsection{Numerical methods}

The simulations are performed in 3D using the Adaptive Mesh Refinement (AMR) code {\ttfamily RAMSES} \citep{teyssier}, incorporating the constrained transport scheme \citep{2004JCoPh.195...17L,fromang,teyssierMHD} and the ambipolar and Ohmic diffusion \citep{masson_nimhd}. 
We use the HLL Riemann solver \citep{HLL1983} with the minmod slope limiter. 
The initial grid is uniform and contains $32^3$ cells (level $5$ of refinement). Cells are then refined to ensure at least 8 points per Jeans length at every location. The stiff equation of state considerably slows down the collapse and prevents the formation of the second Larson core. Considering that the lifetime of the first Larson core is typically a few thousands years \citep[e.g., ][]{2013A&A...557A..90V,2018arXiv180706597B}, we stop our simulation 4 to 5 kyr after the formation of the first Larson core. The maximum resolution reached in the simulations is $\sim 2$ au (level 14 for $M=2$ $M_\odot$ and level 15 for $M=5$ $M_\odot$). The numerical error on angular momentum conservation is of the order of few $10^{51}$ g cm$^2$ s$^{-1}$, which is less than 0.1\% of the total angular momentum and represents less than 10\% of the angular momentum of the disk. The numerical errors are therefore negligible in our analysis.

\subsection{Magnetic resistivities} \label{SecResistivities}

In our monofluid description, the Ohmic and ambipolar diffusion are controlled by their respective resistivity, which are determined by the chemical environment.
We use the table of \citet{2016A&A...592A..18M}, which contains the equilibrium abundances of a reduced chemical network across a wide range of density and temperature. The network includes species relevant to the star formation environment and grains following the MRN size distribution. They account for thermal ionisations, the thermionic emission of grains \citep{deschturner} and the grain evaporation. During the simulation, the resistivities are computed on-the-fly for each cell using the local state variables and tabulated abundances. The method is fairly similar to \citet{UmebayashiNakano1990} and the Non-Ideal magnetohydrodynamics Coefficients and Ionisation Library \citep[NICIL][]{2016PASA...33...41W}, except for the grain evaporation and the thermionic emission that have been included in \citet{2016A&A...592A..18M} and that significantly alter the ionisation at temperatures higher than $750$ K.

\begin{figure}
\begin{center}
\includegraphics[trim=3cm 2cm 0cm 1cm,width=0.49\textwidth]{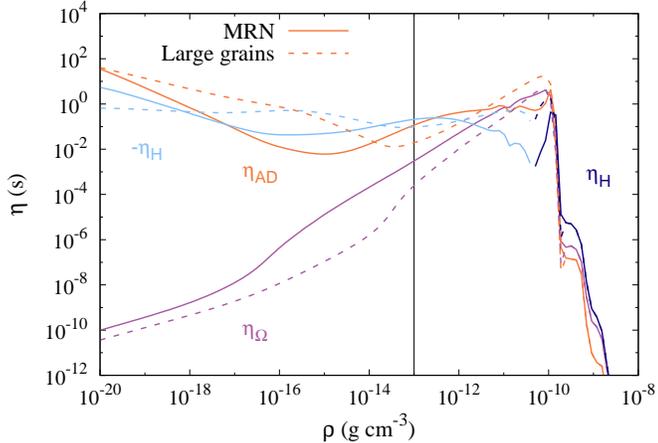}
  \caption{Non-ideal MHD resistivities as a function of density for the MRN distribution (solid lines), and the truncated distribution (dashed lines). Temperature scales with density according to equation \ref{EqEOS}, and, for this figure only, we assumed the following magnetic field prescription $B=1.43 \times 10^{-7} \sqrt{\rho / (\mu_\mathrm{p}m_\mathrm{H})}$ G \citep{LiKrasnopolskyShang}, with $\mu_\mathrm{p}=2.31$ the mean molecular mass and $m_\mathrm{H}=1.67 \times 10^{-24}$ g the mass of one hydrogen atom. The vertical black line represents the first core formation density. $\eta_\mathrm{H}$ represents the Hall resistivity, which can be negative.}
  \label{FigResistivities}
\end{center}
\end{figure}

For four out of eight simulations, we use another table computed in the same manner with a different grain size distribution. In dense cores, at densities typically higher than the ISM, grains smaller than $0.1$ $\mu$m may disappear due to coagulation \citep{2009A&A...502..845O,Guillet2020}. We simulate this coagulation by removing these small grains, while keeping the same dust-to-gas ratio as the initial table, $d=0.0341$\footnote{In the original table, this dust-to-gas ratio was chosen so that the total surface-area of grains matches the uniform distribution of \citet{KunzMouschovias2009}.}, and the same power-law. \cite{2016MNRAS.460.2050Z} show that removing grains smaller than $0.1$ $\mu$m enhances the ambipolar coefficient the most, hence promoting the formation a large disk due to the reduced magnetic braking.
This could be a crucial factor for the AM regulation. The resistivities of both models for the EOS of equation \ref{EqEOS} are displayed in Figure \ref{FigResistivities}. Because the temperature increases rapidly, the grain evaporation and the thermal ionizations occur at $T>1000$ K, near $\rho \approx 10^{-10}$ g cm$^{-3}$, lowering the resistivities near zero. The gas then becomes perfectly coupled with the magnetic field as in the ideal MHD framework. However, purposely, the increasing thermal support prevents the simulations from reaching higher densities.
We qualitatively recover the results of \citet{2016MNRAS.460.2050Z}, with an overall larger ambipolar diffusion and lower Ohmic diffusion coefficients. The main differences is the ambipolar coefficient being smaller in the large-grain cases in the density range $[10^{-14}:10^{-11}]$ g cm$^{-3}$, which could affect the dynamics of the first core and the disk. However, we expect the ambipolar diffusion to work against the magnetic braking mostly in the collapsing envelope, at lower densities. Another discrepancy concerns the sign of the Hall effect, that is not affected in our case. However, both their study and ours do not include it in the simulations. The impact of the Hall term on the evolution of the AM has been examined in great detail in \citet{2018A&A...619A..37M,2019A&A...631A..66M}.
Although included in our simulations, the Ohmic diffusion does not play a major role in the dynamics of the disk at these early stages. Its main effect is the dissipation of the magnetic field in the innermost part of the first Larson core. Since the Ohmic resistivity is sensibly the same with both distributions, we will not quantify its effect in this study.

\section{Results} \label{SecResults}

Disks and outflows form in every simulation. Their properties are reviewed in Sections \ref{Secdisk} and \ref{Secoutflow} respectively, while general properties about the transport of AM are discussed in Section \ref{Secamtransport}. Section \ref{Seciondrift} is dedicated to the analysis of the ion-neutral drift.

\subsection{Disks}\label{Secdisk}

\begin{figure*}
\begin{center}
\includegraphics[trim=0cm 2cm 1cm 0cm, width=0.99\textwidth]{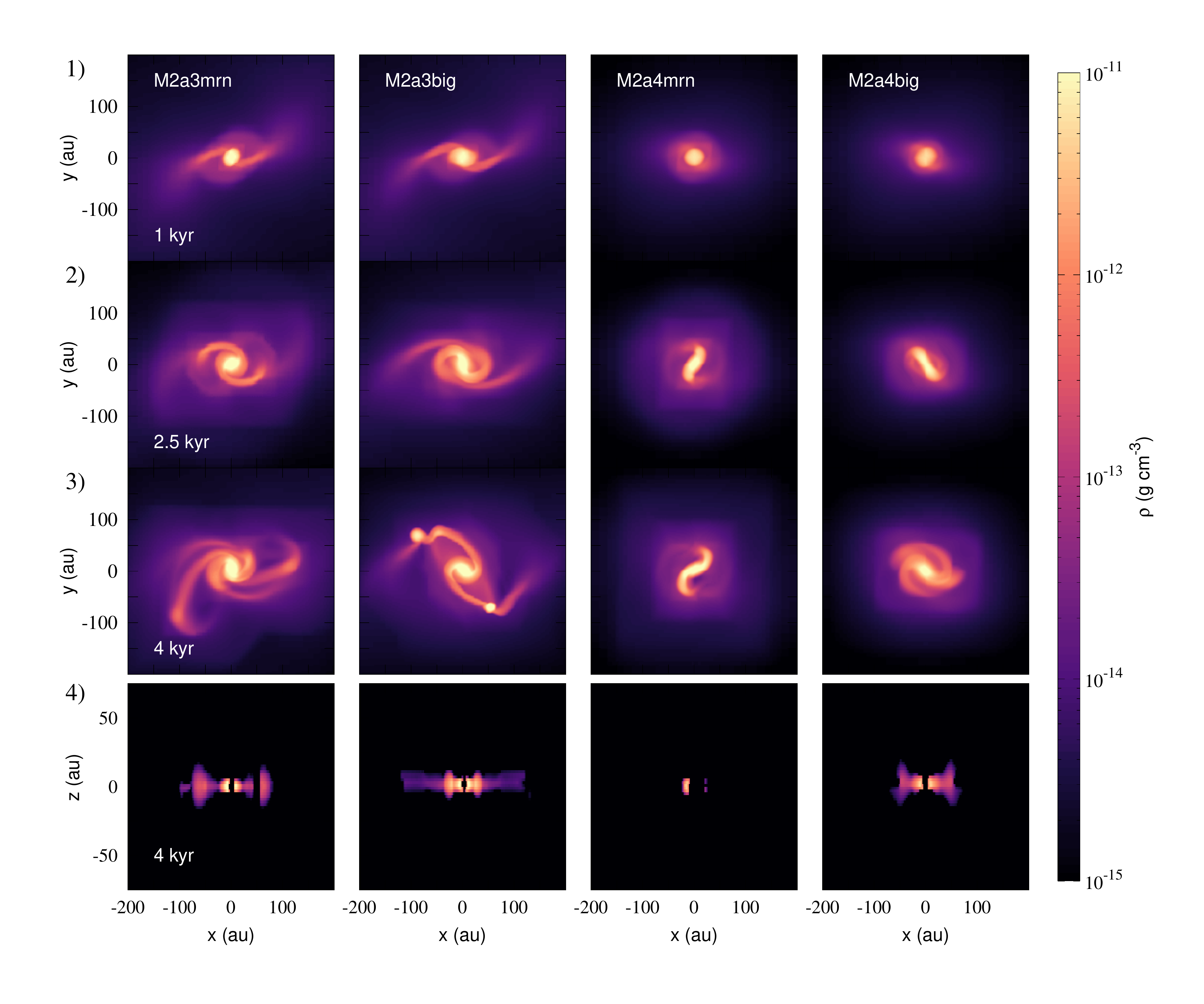}
  \caption{Density maps of the four $M=2$ $M_\odot$ simulations at $t=t_\mathrm{C}+1$ kyr (row 1), $2.5$ kyr (row 2 and 4) and $4$ kyr (row 3). Rows 1 to 3 display slices of the mid-plane, and row 4 shows an edge-on view of the disk at $t-=t_\mathrm{C}+4$ kyr. In row 4, the gas that does not belong to the disk is not shown.}
  \label{Figdiskmapm2}
\end{center}
\end{figure*}

\begin{figure*}
\begin{center}
\includegraphics[trim=0cm 2cm 1cm 0cm, width=0.99\textwidth]{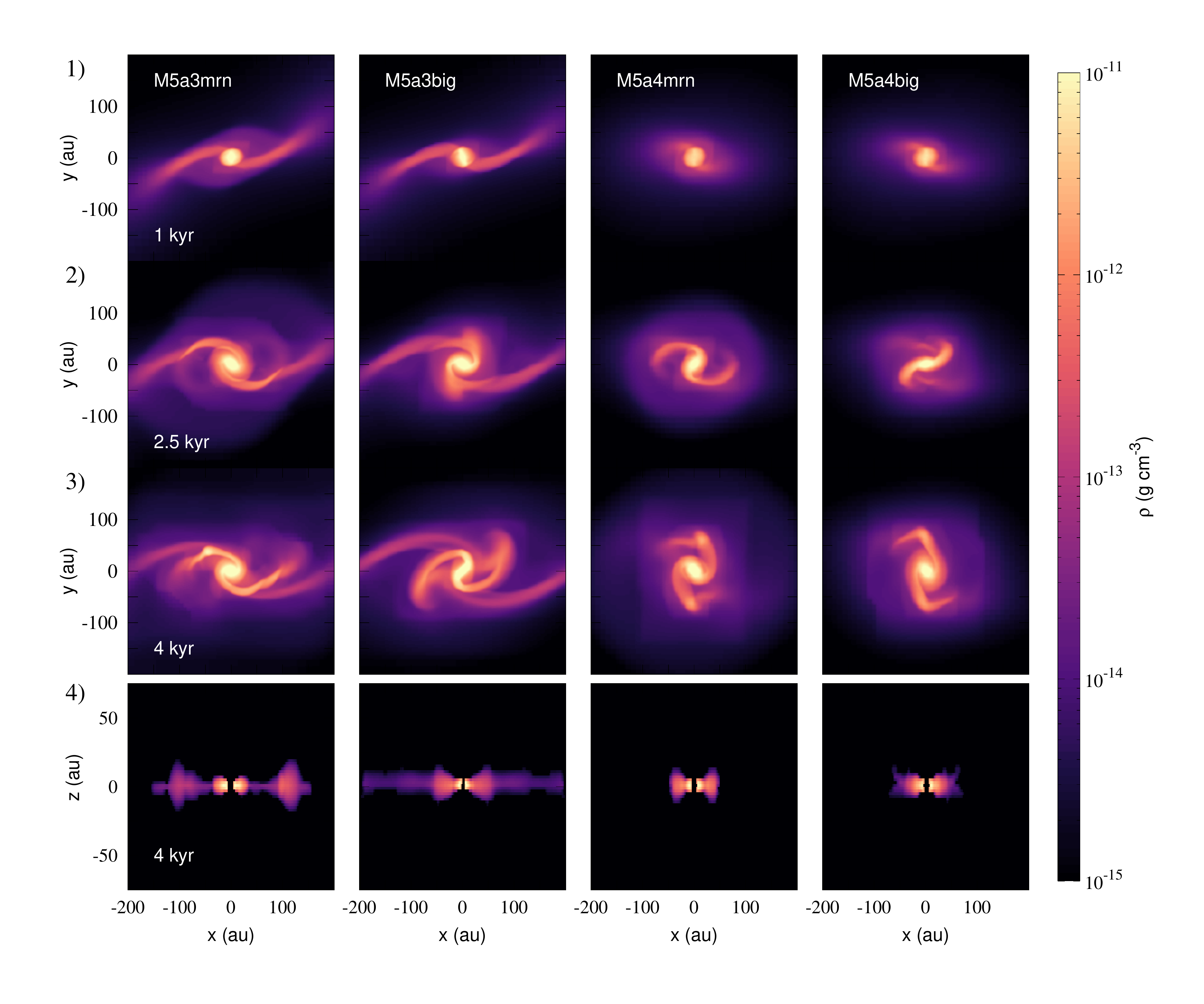}
  \caption{Same as Figure \ref{Figdiskmapm2} for $M=5$ $M_\odot$}
  \label{Figdiskmapm5}
\end{center}
\end{figure*}

\begin{figure*}
\begin{center}
\includegraphics[trim=0cm 2cm 0cm 1cm, width=0.88\textwidth]{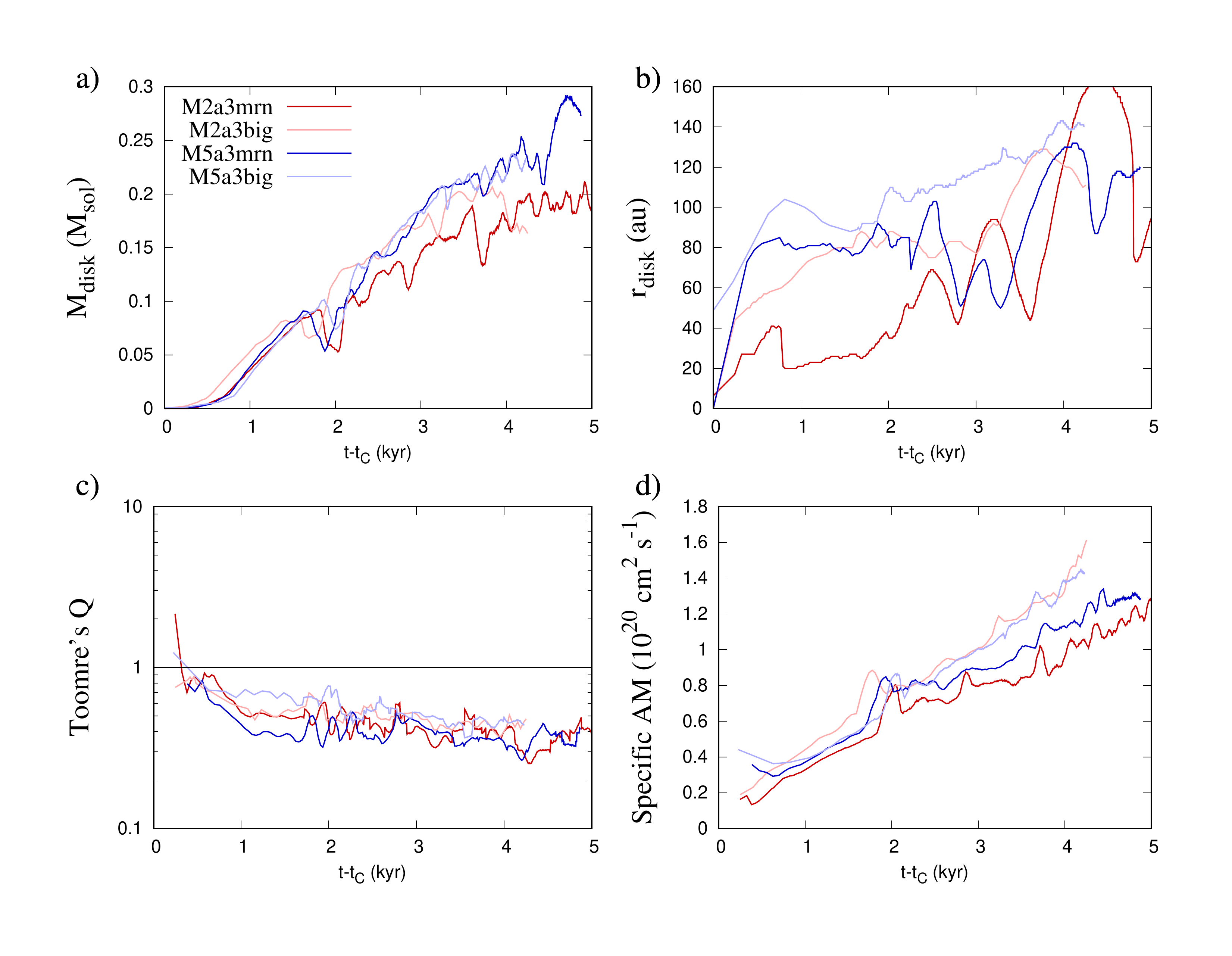}
  \caption{Time evolution of the mass (panel a), radius (panel b) and Toomre's Q (panel c) and specific AM (panel d) of the disk, for $\alpha=0.3$ simulations. Red: $M=2$ $M_\odot$, blue: $M=5$ $M_\odot$. Darker colors represent the standard MRN cases while lighter colors represent the truncated MRN.}
  \label{Figdiskmassrad_a3}
\end{center}
\end{figure*}

\begin{figure*}
\begin{center}
\includegraphics[trim=0cm 2cm 0cm 1cm, width=0.88\textwidth]{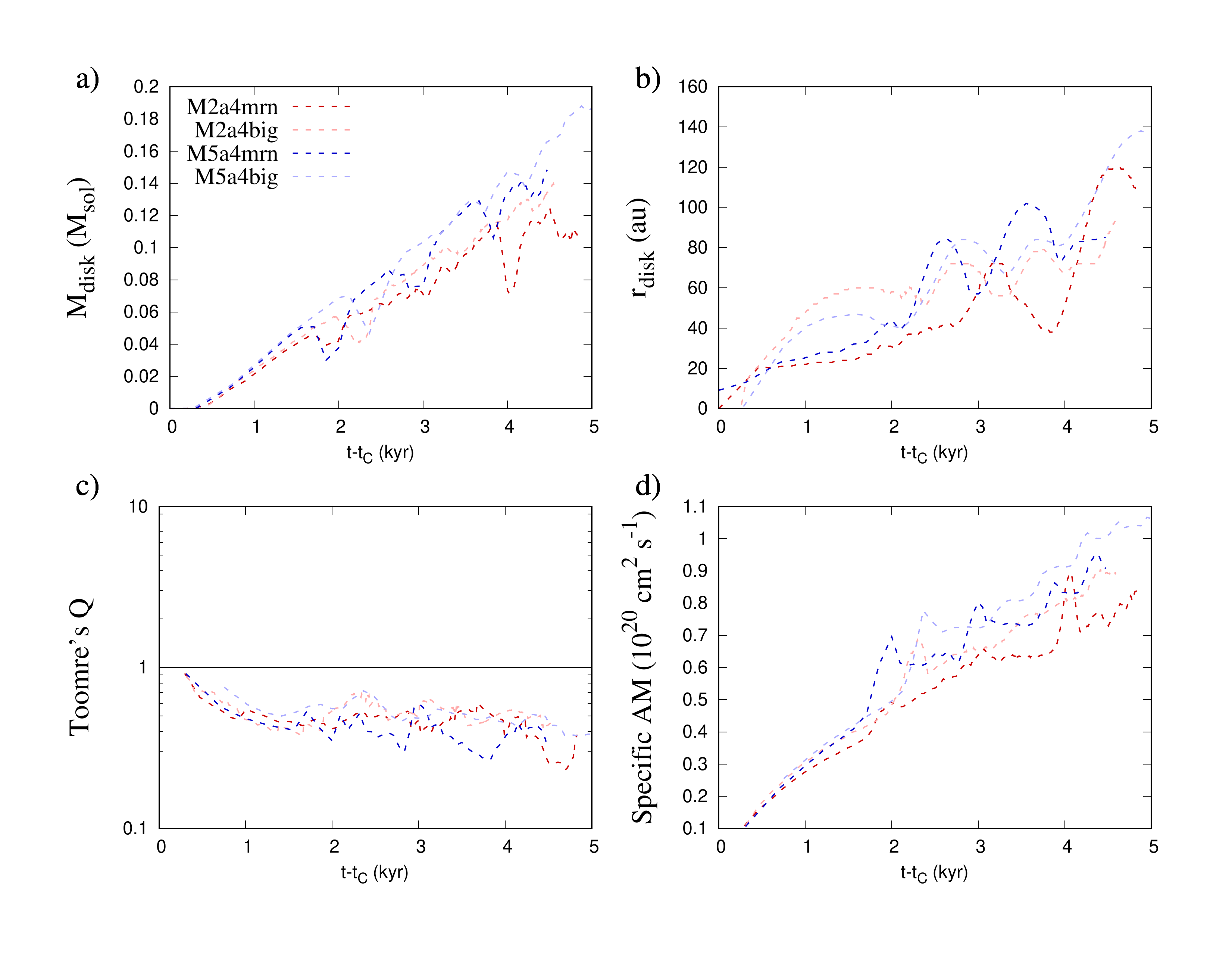}
  \caption{Same as Figure \ref{Figdiskmassrad_a3} for $\alpha=0.4$.}
  \label{Figdiskmassrad_a4}
\end{center}
\end{figure*}

Figures \ref{Figdiskmapm2} and \ref{Figdiskmapm5} display density maps of the mid-plane for $M=2$ $M_\odot$ and $M=5$ $M_\odot$ respectively, at $t=t_\mathrm{C} + 1,2.5$ and $4$ kyr (rows 1 to 3), and an edge-on view of the disk (row 4) at $t=t_\mathrm{C}+4$ kyr. In our analysis, the mid-plane has a thickness of one cell. The disk selection criteria are described in Section \ref{Secinit}. The resolution in the disk ranges from 2 au for the inner 75 au, to 8 au for the outer parts. The formation of spiral arms, that occurs in all the simulations, is triggered by the initial azimuthal perturbation, and they seem overall more prominent for the 5 solar mass cases and for $\alpha=0.3$. M2a3mrn, M2a3big, M5a3mrn and M5a4mrn show significant signs of fragmentation. Figures \ref{Figdiskmassrad_a3} and \ref{Figdiskmassrad_a4} allow a more quantitative analysis by showing the mass, radius, Toomre's Q and AM of the disks. In table \ref{tabledisk}, we also summarize the properties of the cores and the disks at $t=t_\mathrm{C}+4$ kyr.

We define the disk radius as the cylindrical radius containing $99$\% the disk's mass. Toomre's Q parameter is often used to quantify the stability of a disk against fragmentation, given by the following formula
\begin{equation}
  Q = \frac{c_\mathrm{s} \Omega_\mathrm{d}}{\pi G \Sigma},
\end{equation}
with $c_\mathrm{s}$ the sound speed in the disk, $\Omega_\mathrm{d}$ the disk average angular velocity and $\Sigma$ the surface density of the disk. $Q<1$ reflects an instability and a possible fragmentation. The AM is computed over the volume of the disk as
\begin{equation}
  \mathbf{L}_\mathrm{disk} = \int_\mathrm{disk} \rho \mathbf{r}\times\mathbf{u}d\mathrm{V}.
\end{equation}
In this study, we always show the norm of the AM vector $\mathbf{L}$, although its z-component is largely dominant.

Panel a of figures \ref{Figdiskmassrad_a3} and \ref{Figdiskmassrad_a4} shows that the disk mass evolution seems to be weakly dependent on the mass of the initial core and the grain size distribution. The low $M_\mathrm{disk}$ and $r_\mathrm{disk}$ of M2a4mrn in Table \ref{tabledisk} are temporary and due to its fragmentation around this time-mark (see the dark red curve around $4$ kyr in the panel a of Figure \ref{Figdiskmassrad_a4}). The higher thermal support of $\alpha=0.4$ simulations reduces the accretion rate, which yields a lower disk mass. The radii of disks with the truncated distribution are overall larger for $\alpha=0.3$, but only moderately for $\alpha=0.4$. In the same manner, the Toomre's Q do not largely differ between the two grain models. The slightly larger value for the truncated MRN may be due to a self-regulation of spiral arms formed by gravitational instability, that transfer AM faster when they are more unstable. The difference is also clear when looking at the AM evolution. In the larger grains cases, the disks contain an amount of $\sim 1-2 \times 10^{19}$ cm$^2$ s$^{-1}$ more specific AM than their MRN cases counter-part, especially after $2.5$ kyr, which is an increase of 10\% to 20\%. The trend is the same when looking at the normal AM.
The Toomre's Q of all disks are lower than 1 after $2$ kyr, which is reflected by the formation of large spiral arms ($\sim 100$ au length). This low value is however not sufficient for the fragmentation of spiral arms, as was noted by \citet{2018MNRAS.473.4868Z}. \citet{2016MNRAS.458.3597T} shows that the spiral arm fragmentation occurs when the local Toomre's Q in the arms becomes lower than 0.6. Therefore, the fragmentation of the disk can not happen as soon as $Q<1$, but at later stages as it is the case in our simulations. The only run showing no fragmentation by $t=t_\mathrm{C}+4$ kyr is M2a4big, which is reflected by the largest Toomre's Q at $t=t_\mathrm{C}+4$ kyr.
We observe that a larger $\alpha$ provides smaller and less-massive disks with a lower AM. The Toomre's Q is larger and takes more time to reach values below 1. This behavior is mostly a consequence of the lower accretion rate due to the larger free-fall time.

\begin{deluxetable*}{lcccccc}[t]
  \caption{Summary of the properties of the first cores and disks at $t=t_\mathrm{C}+4$ kyr. Mass of the first Larson core, mass of the disk, radius of the disk, Toomre's Q of the disk, mean accretion rate, presence of fragmentation.}
  \label{tabledisk}
  \tablehead{\colhead{Run}          & \colhead{$M_\mathrm{C}$ ($M_\odot$)} & \colhead{$M_\mathrm{disk}$ ($M_\odot$)} & \colhead{$r_\mathrm{disk}$} & \colhead{Toomre's Q} & \colhead{$\dot{M}_\mathrm{acc}$}  & \colhead{Fragmentation}}
\startdata
  M2a3mrn       & $0.30$                     & $0.17$                        &  $123$      & $0.40$    & $7.4\times 10^{-5}$    &   Yes         \\
  M2a3big       & $0.33$                     & $0.19$                        &  $121$      & $0.40$    & $8.2\times 10^{-5}$    &   Yes         \\
  M2a4mrn       & $0.19$                     & $0.07$                        &  $ 52$      & $0.38$    & $4.7\times 10^{-5}$    &   Yes         \\
  M2a4big       & $0.18$                     & $0.13$                        &  $ 71$      & $0.52$    & $4.5\times 10^{-5}$    &   No          \\
  M5a3mrn       & $0.34$                     & $0.22$                        &  $129$      & $0.35$    & $8.5\times 10^{-5}$    &   Yes         \\
  M5a3big       & $0.34$                     & $0.21$                        &  $143$      & $0.40$    & $8.4\times 10^{-5}$    &   Yes         \\
  M5a4mrn       & $0.20$                     & $0.13$                        &  $ 77$      & $0.38$    & $5.1\times 10^{-5}$    &   Yes         \\
  M5a4big       & $0.20$                     & $0.15$                        &  $ 82$      & $0.46$    & $5.0\times 10^{-5}$    &   Yes         \\
\enddata
\end{deluxetable*}

\begin{figure}
\begin{center}
\includegraphics[trim=2cm 1.5cm 0cm 2cm, width=0.48\textwidth]{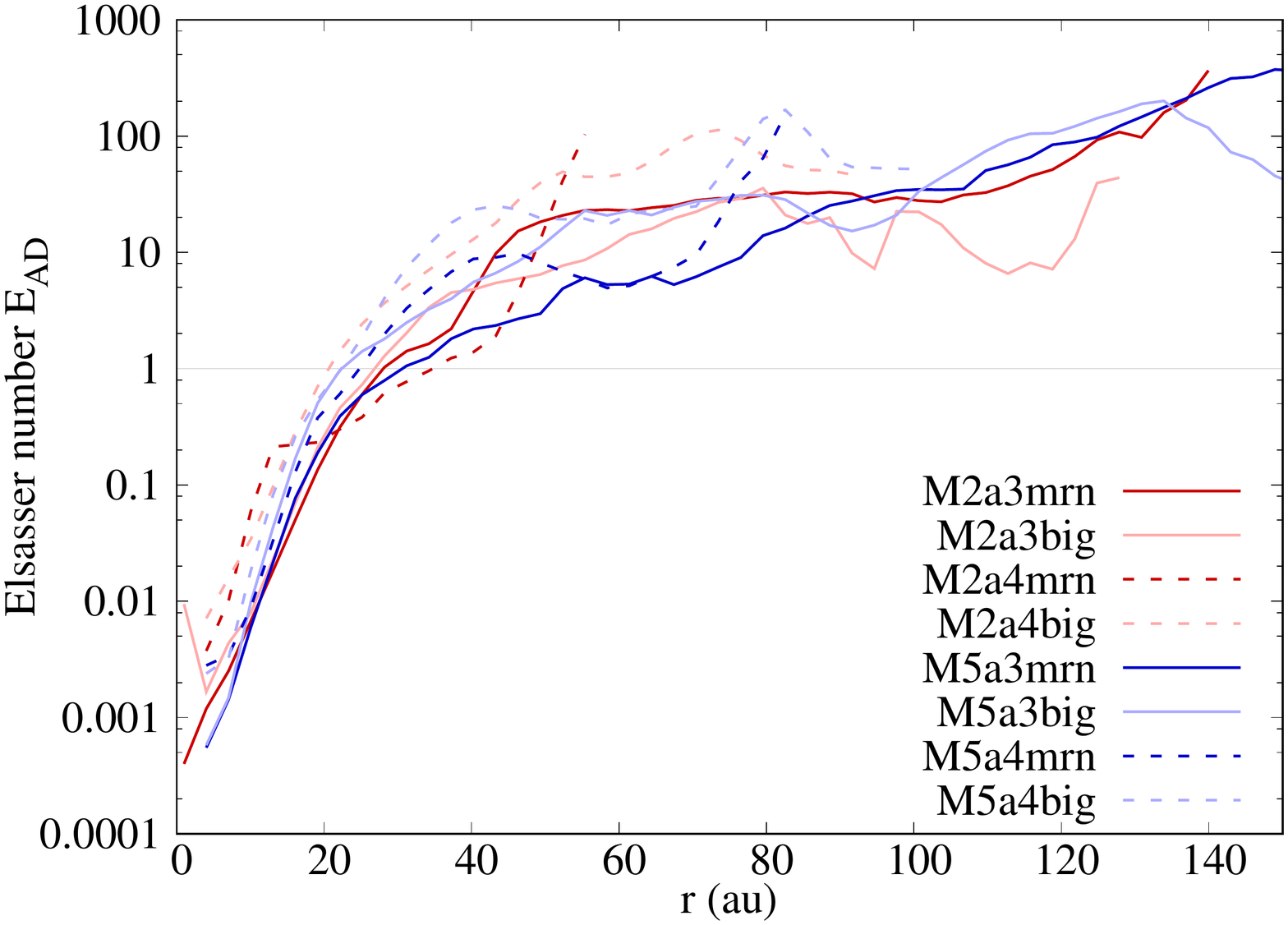}
  \caption{Radial profiles of the ambipolar Elsasser number (Am) in the disk for the eight simulations at $t=t_\mathrm{C}+2.5$ kyr. The quantities have been density-averaged over the azimuthal direction.}
  \label{Figelsasser}
\end{center}
\end{figure}

To quantify the influence of the ambipolar diffusion in the disk, we calculate the Elsasser number, which is defined as

\begin{equation}
  \mathrm{E_{AD}} = \frac{4\pi u_\mathrm{A}^2}{c^2\eta_\mathrm{AD}\Omega},
\end{equation}
with $u_\mathrm{A}=B/\sqrt{\rho}$ the Alfven speed, and $\Omega$ the angular velocity of the fluid. E$_\mathrm{AD}$ $< 1$ means that the ambipolar diffusion has a significant impact over the dynamics of the disk. Figure \ref{Figelsasser} displays the radial profiles of Am in the eight different disks, in which we have performed an azimuthal density-weighted average. 
Only the innermost $20$ to $40$ au parts of the disks show E$_\mathrm{AD}$ $<1$, down to $10^{-3}$. In the outer regions, E$_\mathrm{AD}$ rises up to $10$ to $100$, indicating a less efficient decoupling between the magnetic field and the gas compared to the central domain. As a consequence of the higher ambipolar resistivity of the reference MRN distribution in the density range $[10^{-14}:10^{-11}]$ g cm$^{-3}$, the disks with the non-truncated MRN distribution have larger E$_\mathrm{AD}$ $<1$ regions than their truncated-MRN counter-part, up to a factor 2 for M2a3mrn and M2a3big. The truncated MRN disks contains however a larger angular momentum (see panel d of figures \ref{Figdiskmassrad_a3} and \ref{Figdiskmassrad_a4}), meaning that this difference is not significant compared to the influence of the ambipolar diffusion in the outer parts of the disk and in the envelope, although with a higher Elsasser coefficient.

\begin{figure*}
\begin{center}
\includegraphics[trim=1cm 1cm 1cm 0cm,width=0.99\textwidth]{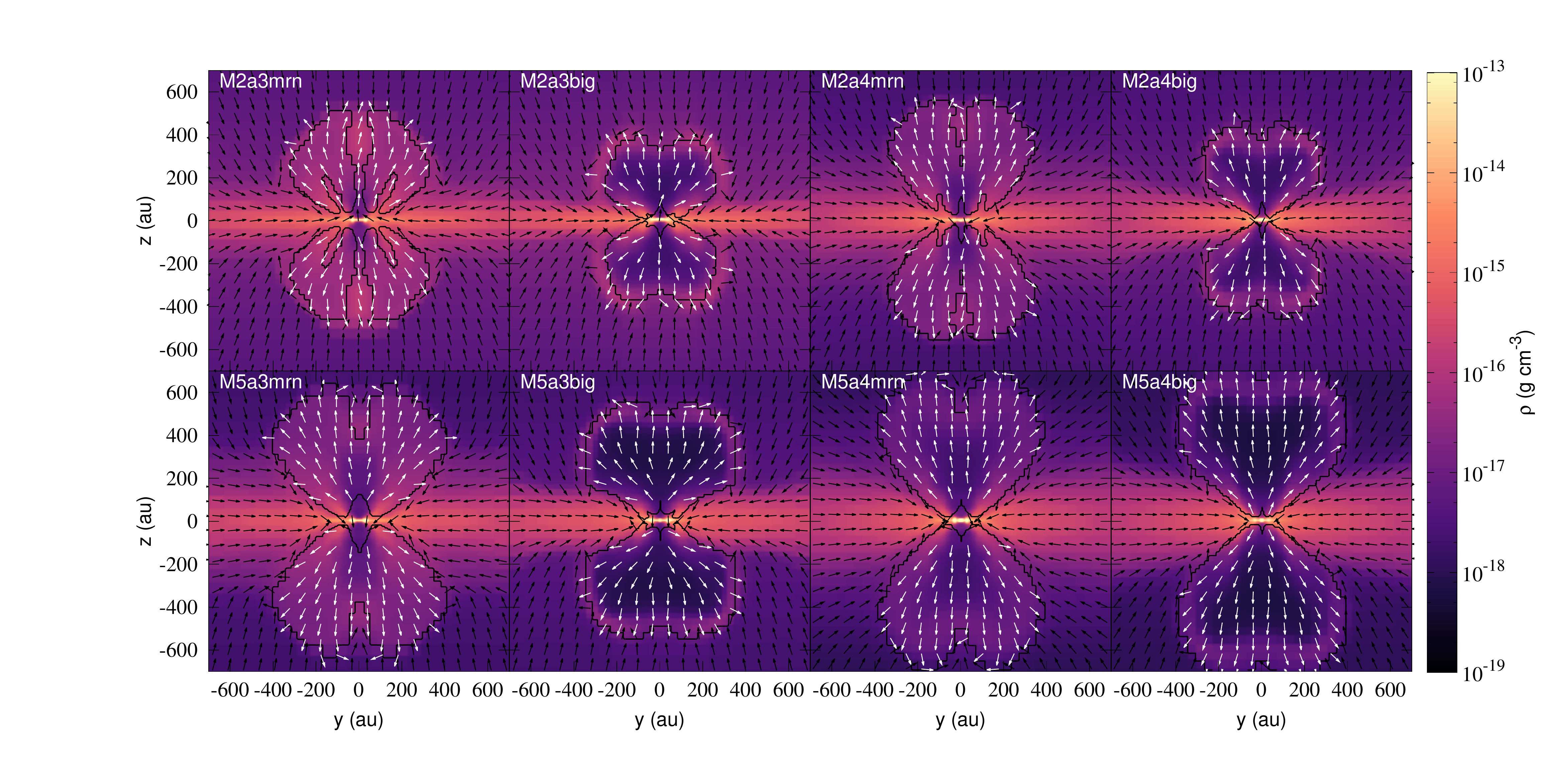}
  \caption{Density slices along the plane x=0 for all simulations at $t=t_\mathrm{C}+2.5$ kyr. The black lines represent the contour of the outflows and arrows indicate the fluid velocity. Infalling material is marked with black arrows while the outflow is marked with white arrows. Top: M=$2$ $M_\odot$, bottom: M=$5$ $M_\odot$. From left to right: M*a3mrn, M*a3big, M*a4mrn, M*a4big.}
  \label{Figoutflowmap}
\end{center}
\end{figure*}

\subsection{Outflows}\label{Secoutflow}

\begin{figure}
\begin{center}
\includegraphics[trim=2cm 1cm 0.5cm 1cm,width=0.49\textwidth]{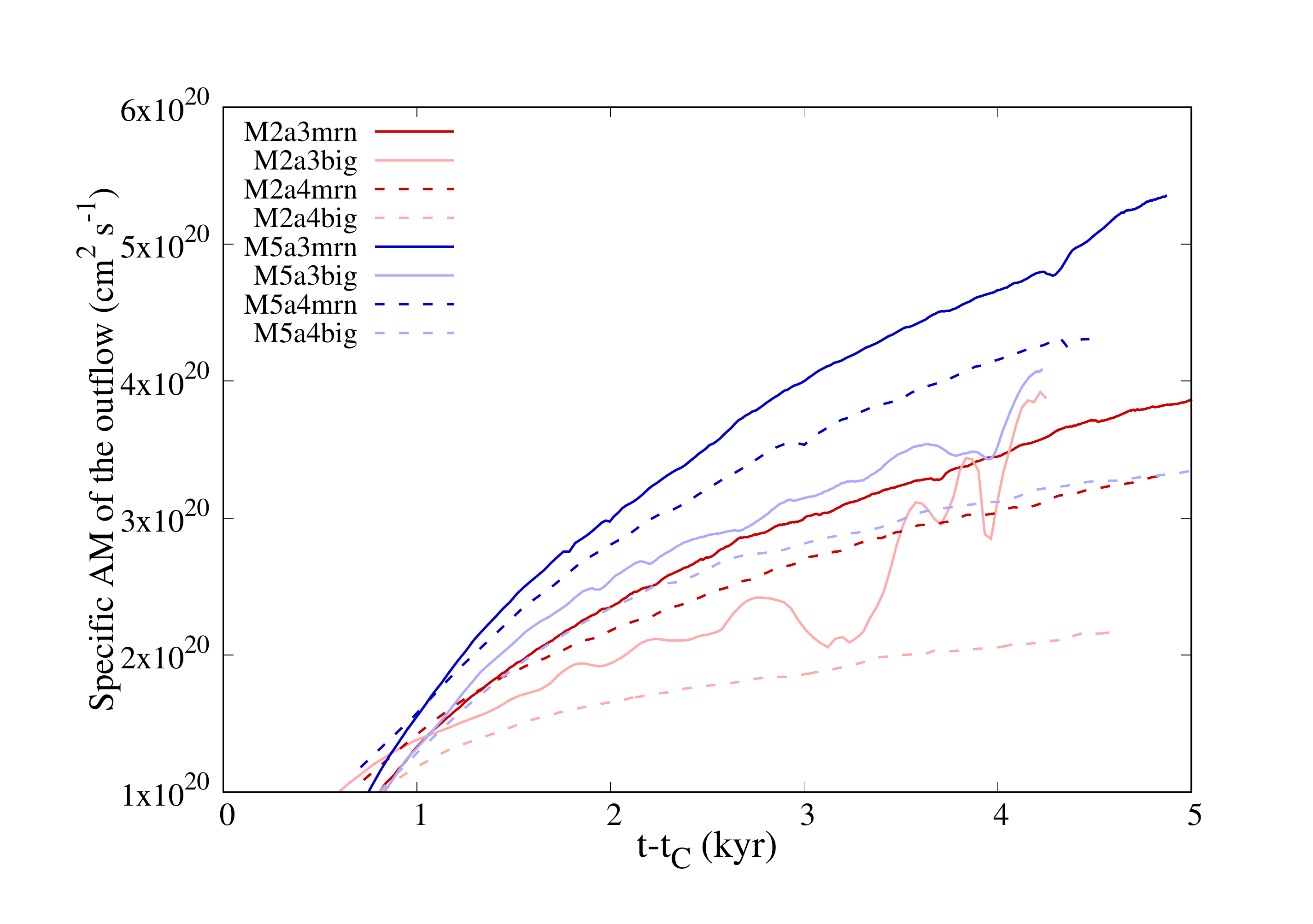}
\includegraphics[trim=2cm 1cm 0cm 1cm,width=0.49\textwidth]{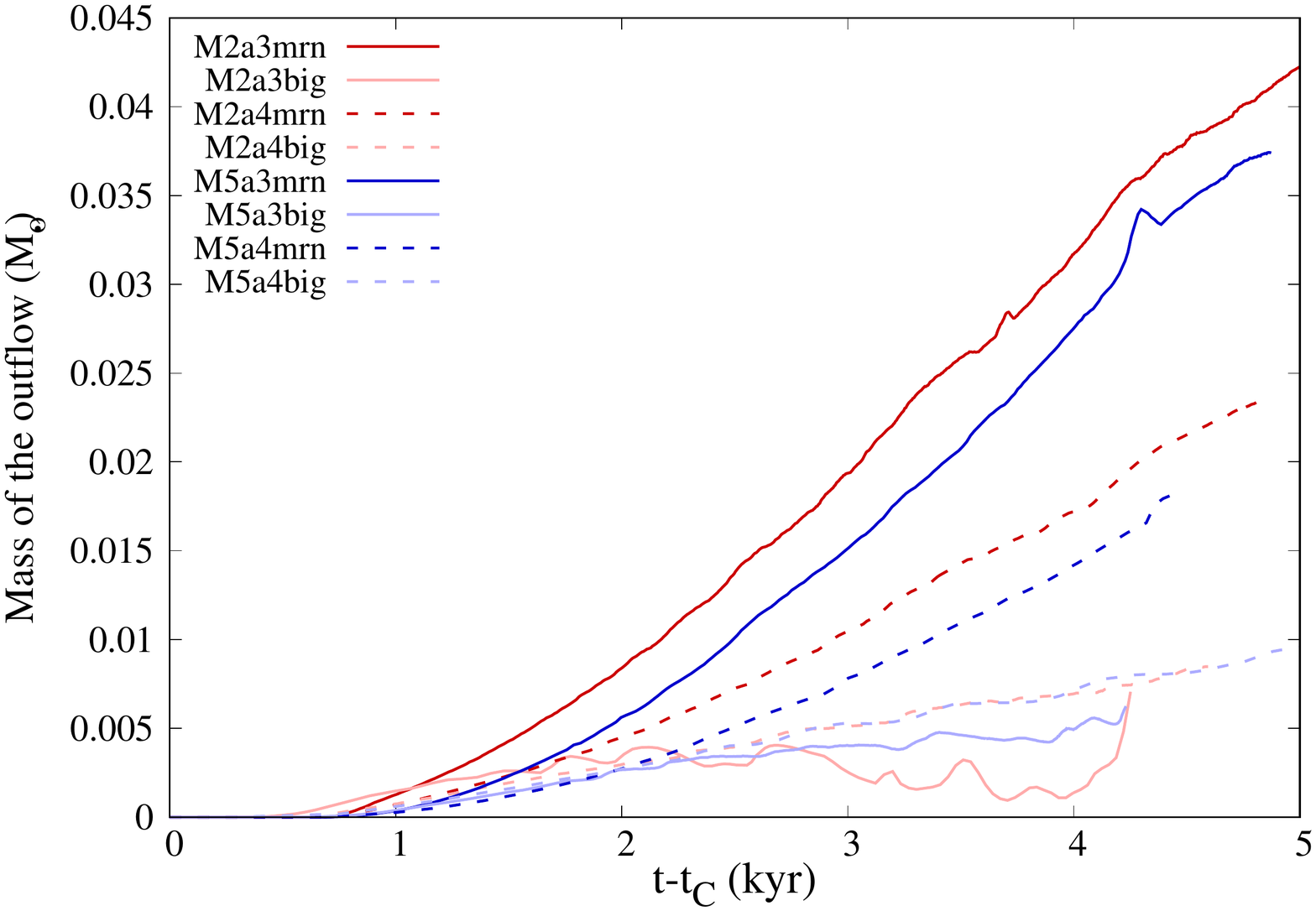}
  \caption{Time evolution of the specific AM (top panel) and the mass (bottom panel)} of the outflow. The curves are the same as in Figure \ref{Figdiskmassrad_a3} and \ref{Figdiskmassrad_a4}.
  \label{Figangoutflows}
\end{center}
\end{figure}

All simulations produce bipolar outflows that start within 1 kyr after the first core formation. Figure \ref{Figoutflowmap} shows density maps in the outflow region at $t-t_\mathrm{C}=2.5$ kyr. The outflows have similar heart-like structures. It appears that the truncated MRN cases produce outflows with an overall slower expansion and a lower density of a factor $\sim$ 20 in the cavity ($\approx 5\times 10^{-17}$ g cm$^{-3}$ vs $\approx 2\times 10^{-18}$ g cm$^{-3}$) than the standard distribution. They show a clear cavity with dense borders, while cavity only starts to form at $2.5$ kyr with the MRN distribution. The weakness of the truncated MRN outflows are further confirmed when looking at their specific AM and their mass, as shown in Figure \ref{Figangoutflows}. With the truncated MRN distribution, the specific AM is 25\% to 33\% lower than with the standard MRN. In addition, the bottom panel of the figure shows that outflows of the truncated MRN distribution eject significantly less gas. While the mass outflow rate is $\approx 10^{-5}$ M$_\odot$ yr$^{-1}$ for the normal MRN with $\alpha=0.3$, and $\approx 6\times 10^{-6}$ M$_\odot$ yr$^{-1}$ with $\alpha=0,4$, all four truncated MRN cases have mass loss rates $\lesssim 2.5 \times 10^{-6}$ M$_\odot$ yr$^{-1}$.

The cases with $\alpha=0.4$ also eject less AM for the normal distribution, but a similar amount than $\alpha=0.3$ with the truncated MRN. However these quantities do not depend on the dense core's mass. These observations remain qualitatively unaltered if considering the specific AM instead.

\begin{figure}
\begin{center}
\includegraphics[trim=9cm 3cm 3cm 3cm,width=0.21\textwidth]{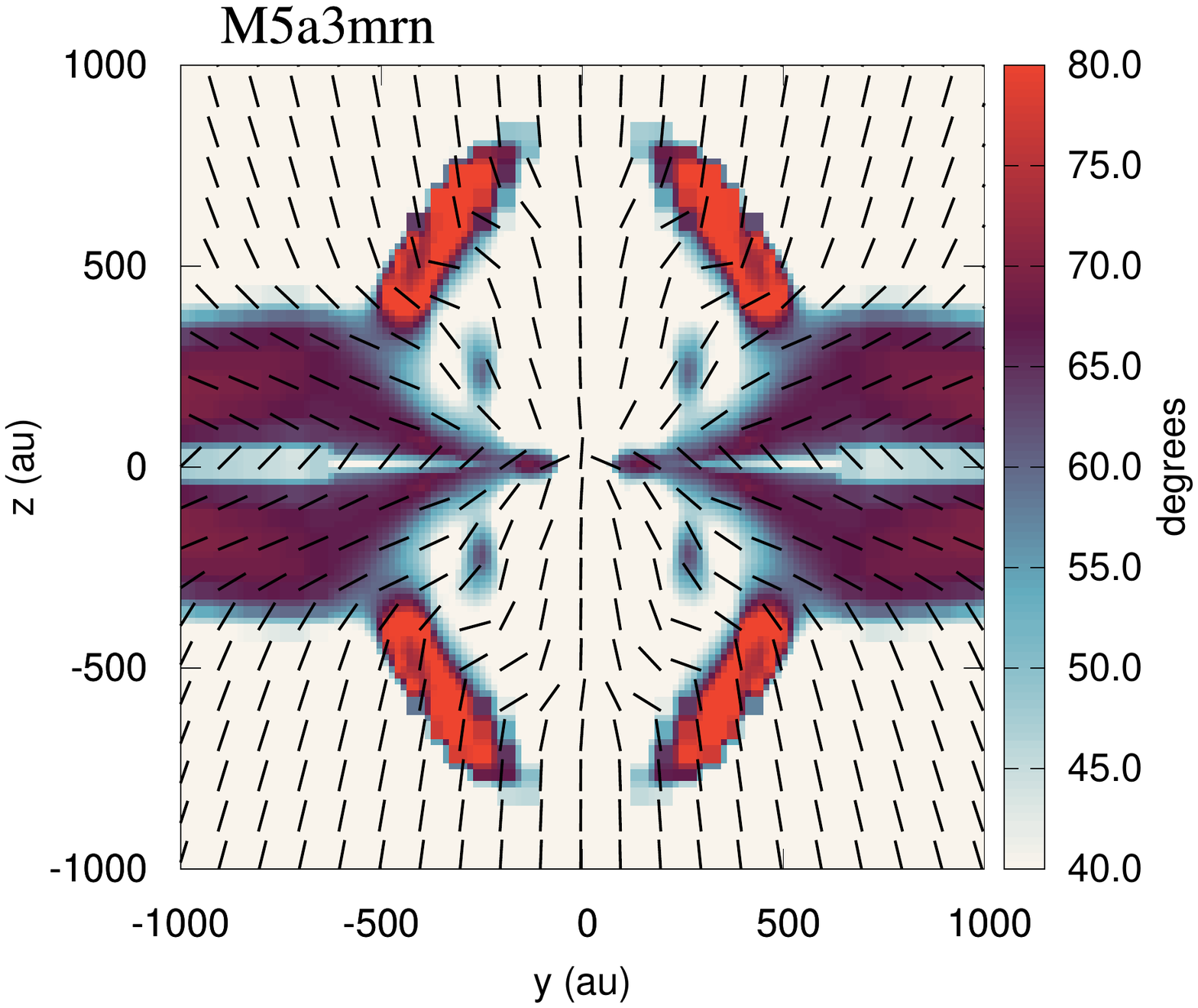}
\includegraphics[trim=6cm 3cm 6cm 3cm,width=0.21\textwidth]{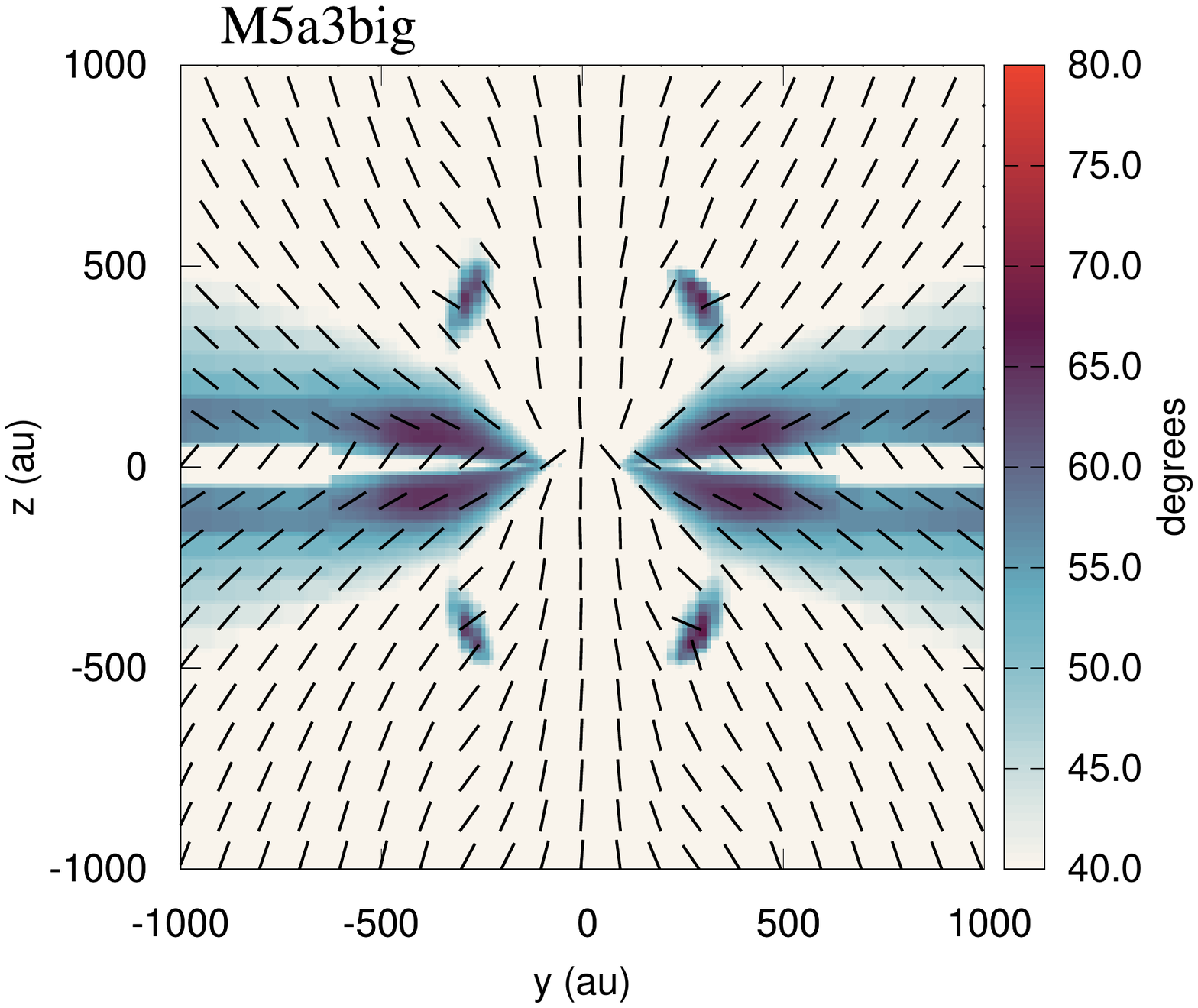}
\includegraphics[trim=9cm 3cm 3cm 3cm,width=0.21\textwidth]{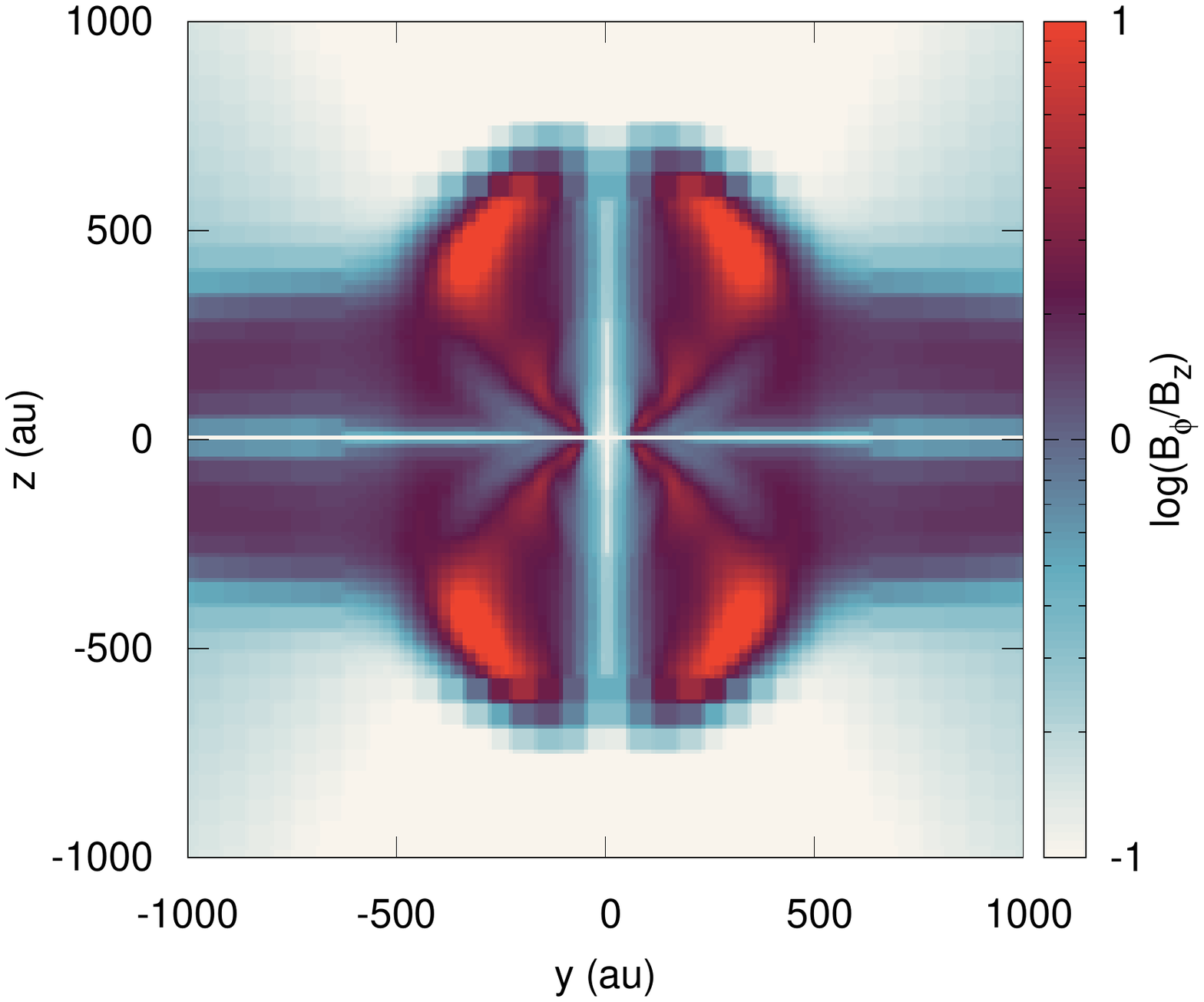}
\includegraphics[trim=6cm 3cm 6cm 3cm,width=0.21\textwidth]{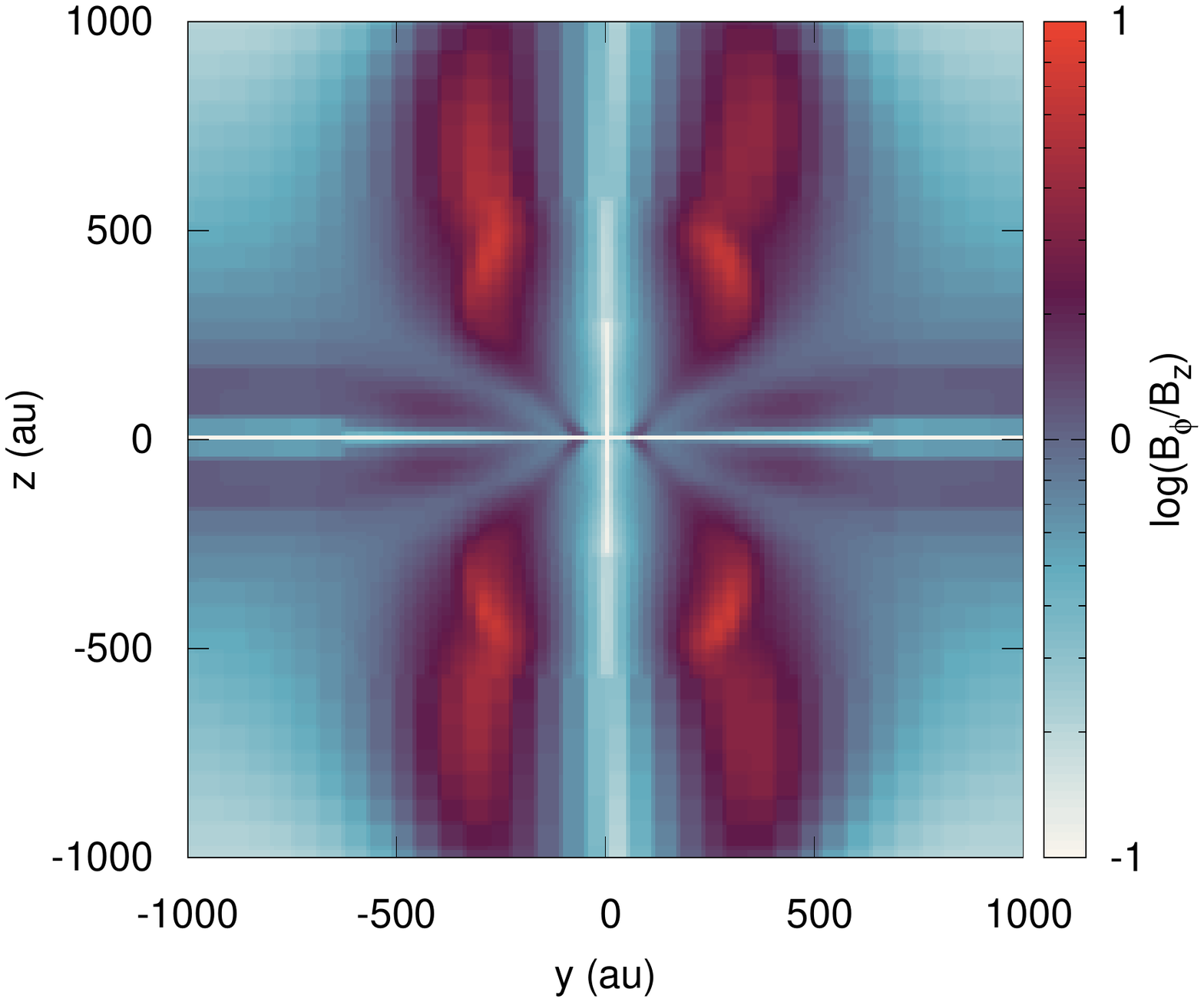}
  \caption{Top panels: Angle between the local magnetic field and the initial magnetic field (vertical) in the y-z plane, at $t=t_\mathrm{C}+2.5$ kyr, calculated as $\mathrm{atan}(B_r/B_z)$. The segments represent the direction of the magnetic field. Bottom panels: log($B_\phi/B_z$). Left column is M5a3mrn and right column is M5a3big.}
  \label{Figangleb}
\end{center}
\end{figure}

In our simulations, the outflows are launched by \citet{1982MNRAS.199..883B} mechanism. The magnetic field lines are bent by the drag of the collapsing gas, producing the well-known hourglass-shape. If the angle is high enough, the centrifugal acceleration added to the Lorentz force may eject the gas along the field lines. In the default MRN cases, the field lines can be bent up to an angle of $75\degree$ with the z-axis near the central region, which is a $15\degree$  slope for the gas acceleration in the outward radial-direction. However, with the truncated MRN, the higher ambipolar coefficient in the envelope reduces the accretion of the magnetic field due to a weaker coupling with the gas, resulting in a maximum angle of $65\degree$. These inclination angles of the magnetic field, calculated as $\mathrm{atan}(B_r/B_z)$, are displayed in Figure \ref{Figangleb} for both M5a3mrn and M5a3big simulations, 2.5 kyr after the formation of the first Larson core, in the plane $x=0$. The pseudo-disk and the contour of the outflow appear clearly. Figure \ref{Figangleb} also include maps of $B_\phi/B_z$, which shows that the relative strength of $B_\phi$ is overall higher with the truncated distribution. We discuss this point more in details in Section \ref{Seciondrift}. Another consequence of the decoupling is the reduced magnetic field strength with the truncated distribution, twice as low as with the default distribution in the vicinity of the outflow launching region. Both these effects reduce the magneto-rotational acceleration of the gas. However, while it is difficult to accurately measure the opening angle of the outflows, we estimate both numerically and visually that the angle is systematically larger for the truncated-MRN simulations by 5 to 20 degrees, which is somewhat counter-intuitive. This is because the outflow is confined by the magnetic pressure of the surrounding gas. The lower ambipolar coefficient in the standard MRN simulations allows for more magnetic field accretion, and therefore an increased magnetic pressure that reduces the radial extension of the outflow.

\subsection{Angular momentum transport}\label{Secamtransport}

\subsubsection{The magnetic braking}

In this section, we look at the AM regulation in the various simulations. There are two main AM transport mechanisms at play here, that are the advection by the gas (by the gravitational collapse or the outflow), and magnetic braking. To compute the cumulative losses of angular momentum by magnetic braking up to a given time $t$, we integrate the magnetic torque over the disk or the pseudo-disk

\begin{equation}
  L_\mathrm{mag}=\int_{t_\mathrm{c}}^t \int_\mathrm{V} r(\mathbf{J}\times\mathbf{B})_\phi d\mathrm{V}  dt.
\end{equation}
where V represents the volume of integration. The criterion to define the pseudo-disk is simply $\rho > 10^{-15}$ g cm$^{-3}$, excluding the disk and the first core as defined in the previous sections.

\begin{figure}
\begin{center}
\includegraphics[trim=2cm 2.5cm 0cm 1cm,width=0.49\textwidth]{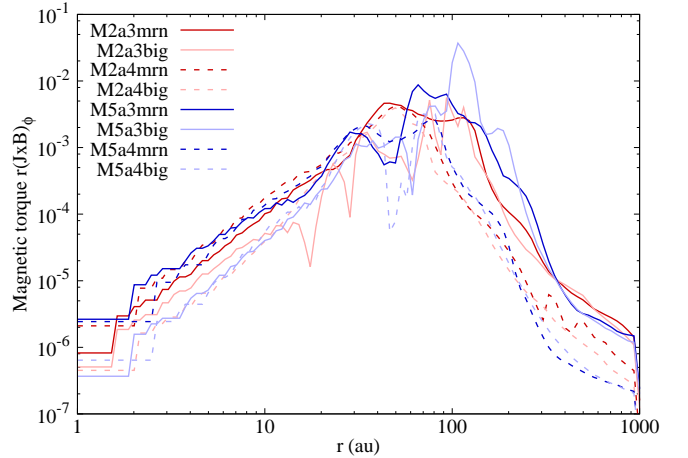}
  \caption{Azimuthal average of the magnetic torque $r \left( \mathbf{J}\times \mathbf{B}\right)_{\varphi}$ in the mid-plane as a function of the cylindrical radius.}
  \label{Figmagbrakprofile}
\end{center}
\end{figure}

The profile of the magnetic torque $r \left( \mathbf{J}\times \mathbf{B}\right)_{\varphi}$ in the mid-plane at $t=t_\mathrm{C}+4$ kyr is displayed in Figure \ref{Figmagbrakprofile}. The evolutions are similar between the eight cases, with a dispersion across one order of magnitude. The magnetic torque increases as $r^2$ until $r\approx 100$ au, meaning $\left( \mathbf{J}\times \mathbf{B}\right)_{\varphi} \propto r$. Then, the torque decreases rapidly as $r^{-4}$. The location of the peak indicates that the magnetic braking is mostly active in the outer regions of the disk (here $30 \lesssim r \lesssim 150$ au). This trend is consistent with the low value of the Elsasser number in the central region, indicating a significant action of the ambipolar diffusion against the magnetic braking. Within the inner $20$ au of the disk, the magnetic torque is larger for the standard MRN than for the truncated MRN, by a factor 2 to 10. Conversely, $\alpha$ has a limited impact on the magnetic torque within the innermost $40$ au, where the ambipolar diffusion dominates. We observe the opposite phenomenon outside the disk. While the larger $\alpha$ decreases the magnetic torque by one order of magnitude, the grain size-distribution does not alter it. However, we probe here only the mid-plane of the pseudo-disk, in which the ambipolar resistivity is low in both cases. Most of the magnetic braking would take place in outer layers.

\begin{figure}
\begin{center}
\includegraphics[trim=0cm 0cm 0cm 0cm,width=0.4\textwidth]{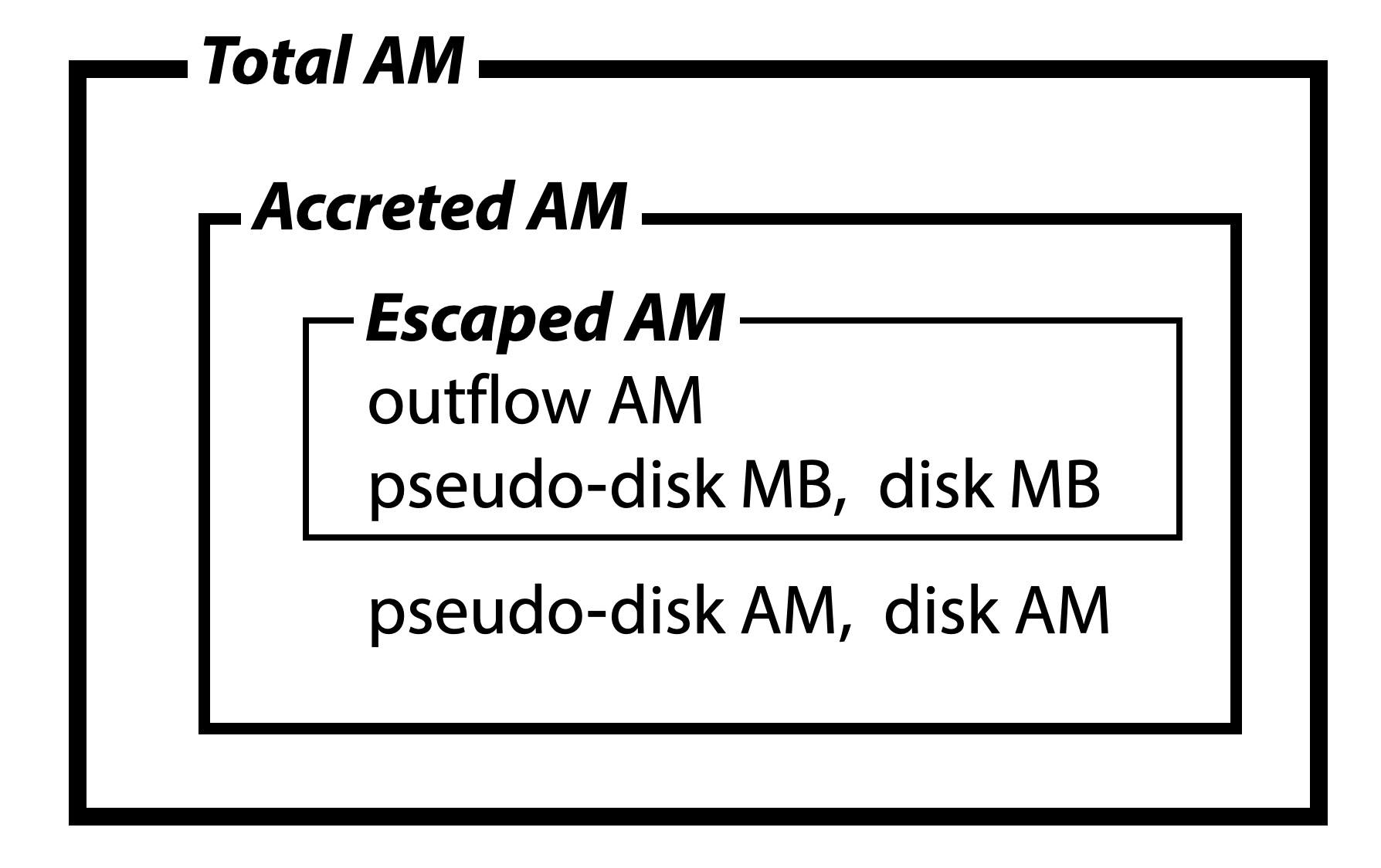}
\caption{Schematic representation of the definitions of the Total AM, the Accreted AM and the Escaped AM.}
\label{Figrefam}
\end{center}
\end{figure}

\begin{figure*}
\begin{center}
\includegraphics[trim=0.5cm 8cm 0.5cm 1.5cm,width=0.99\textwidth]{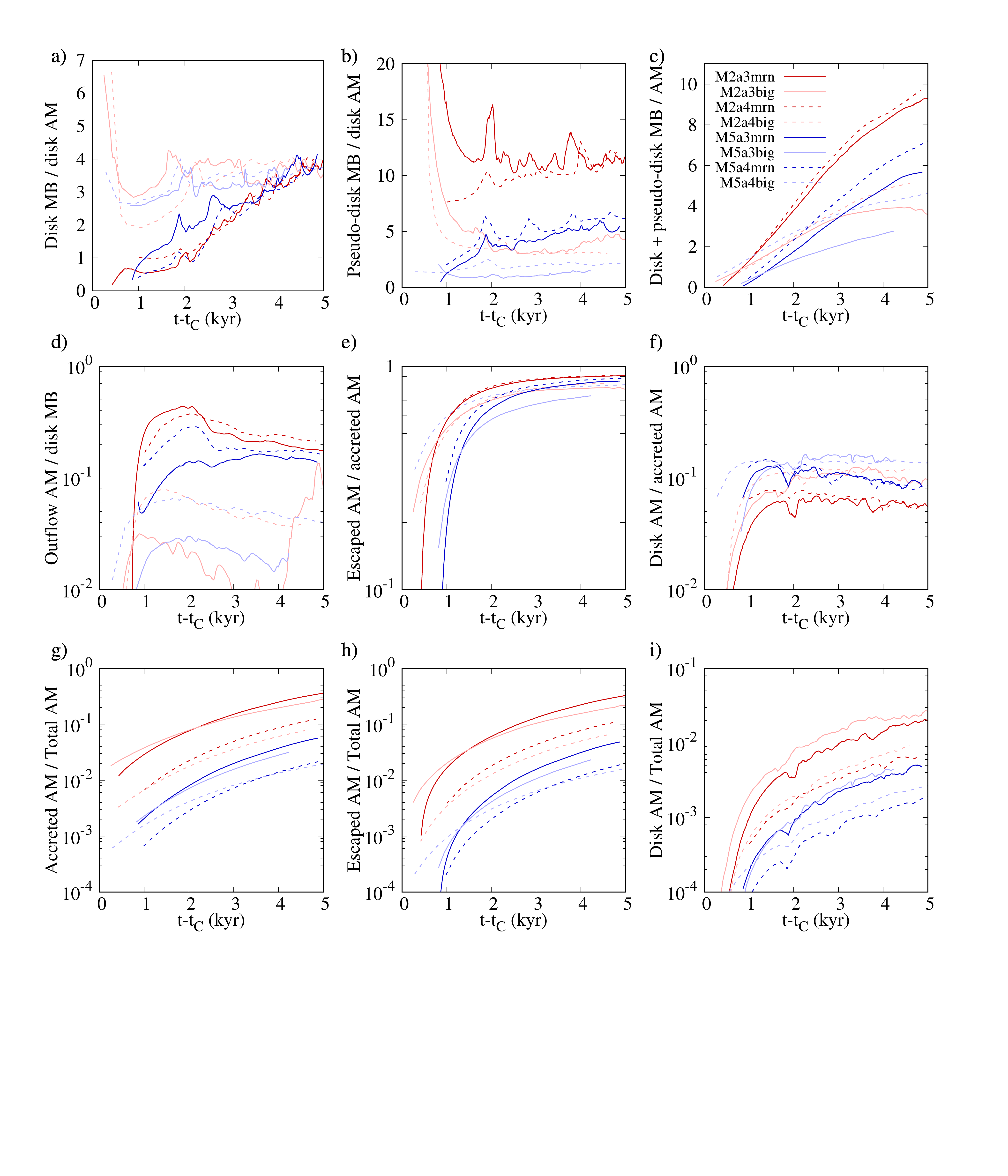}
  \caption{Panel a: ratio between the cumulative magnetic braking in the disk and the AM of the disk. Panel b: ratio between the cumulative magnetic braking in the pseudo-disk (not including the disk) and the AM of the disk. Panel c: Ratio between the cumulative magnetic braking in the disk + pseudo-disk and the AM of the disk + pseudo-disk. Panel d: Ratio between the AM transported by the outflow and the cumulative magnetic braking in the disk. Panel e:  Ratio between the escaped AM and the accreted AM. Panel f: Ratio between the AM of the disk and the accreted AM. Panel g: Ratio between the accreted AM (accreted AM = disk + pseudo-disk + MB in the disk and pseudo-disk + outflow) and the total AM. Panel h:  Ratio between the total escaped AM (MB in the disk and the pseudo-disk + outflow), and the total AM of the dense core. Panel i: Ratio between the disk's AM and the total AM of the dense core.}
  \label{FigBrakdisk}
\end{center}
\end{figure*}

Figure \ref{FigBrakdisk} compares the disk and pseudo-disk AM with the amount extracted by magnetic braking, as well as with the AM taken away by the outflow. Panel a) shows that the magnetic braking has removed three to four times (at $t=t_\mathrm{C}+5$ kyr) the AM left in the disk, meaning that only 20 to 25 percent of the accreted AM remains in the disk. The magnetic braking is more efficient in the disk for the truncated MRN simulations, which is consistent with a slightly lower ambipolar diffusion coefficient (see Fig. \ref{FigResistivities}). However, panel b) shows that the magnetic braking is the most active in the pseudo-disk. With the MRN distribution, more AM is removed by magnetic braking in the pseudo-disk (up to a factor 3 at $t=t_\mathrm{C}+5$ kyr). With the truncated distribution, the magnetic braking is similar in both region. The pseudo-disk is therefore of prime importance when considering the magnetic braking. This is also demonstrated by panel c), which shows the ratio of magnetic braking and AM for the disk and the pseudo-disk combined. Only 10 to 14 percent of the AM remains in this structure for the MRN distribution, and 16 to 25 percent for the truncated MRN cases. 
The magnetic braking is stronger in the lower-mass cores of $M=2$ M$_\odot$ simulations than in $M=5$ M$_\odot$ simulations. Conversely, the $\alpha$ parameter does not change sensitively the amount of AM removed by magnetic braking.
Panel d) displays the ratio between the AM taken away by the outflow and the magnetic braking. The magnetic braking largely dominates the outflow by a factor of two to ten with the MRN distribution, and ten to a hundred with the truncated MRN. In this latter case, both the magnetic braking and outflows remove significantly less AM than in the MRN distribution, which further confirms the results of \citet{2016MNRAS.460.2050Z}.
Panel e) presents the ratio of escaped AM (MB in the disk and pseudo-disk + outflow) to the accreted AM (that comprises the AM of the disk, the pseudo-disk, the outflow and the amount of AM removed by MB in the disk+pseudo-disk) (See Figure \ref{Figrefam} for a schematic view of the different considered AM). It shows that $80-90\%$ of the accreted angular momentum escapes by the outflow or the MB. Panel f) shows the ratio between the AM of the disk and the accreted AM. This ratio remains constant at about $10\%$ throughout the life first Larson core, complementary to the $\approx 90\%$ ratio of the escaped AM.  Panels g), h) and i) show the accreted AM, the escaped AM and the AM of the disk compared to the total AM of the core. By $t=t_\mathrm{C}+4$ kyr, 0.1\% to 3\% of the total AM has made its way and stayed into the disk, concentrated in $5 - 10\%$ of the total mass. As expected, the ratio is systematically larger for the truncated MRN cases. Panel h) indicates that 1\% to 30\% of the total AM have been removed by MB or the outflow by $t=t_\mathrm{C}=4$ kyr, this amount being larger for the normal MRN. All the 2 M$_\odot$ clouds have accreted a larger fraction of their total AM than the 5 M$_\odot$ cases, and thus concentrate a larger fraction of the total AM into their disk, whilst experiencing a larger loss of AM.

\subsubsection{AM in the outflow}

\begin{figure}
\begin{center}
\includegraphics[trim=5cm 3cm 0.5cm 3cm,width=0.49\textwidth]{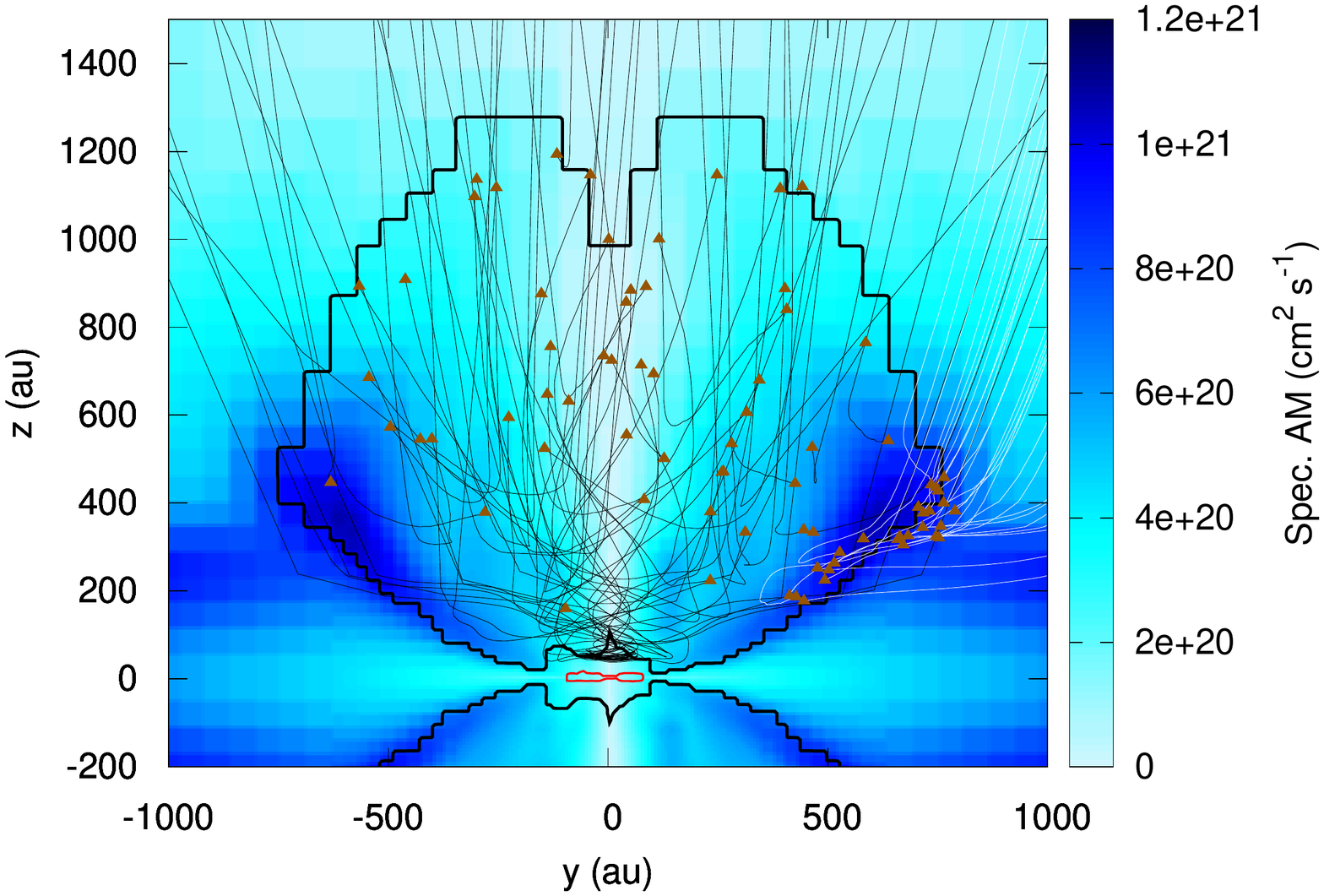}
\includegraphics[trim=6cm 2cm 2cm 3cm,width=0.49\textwidth]{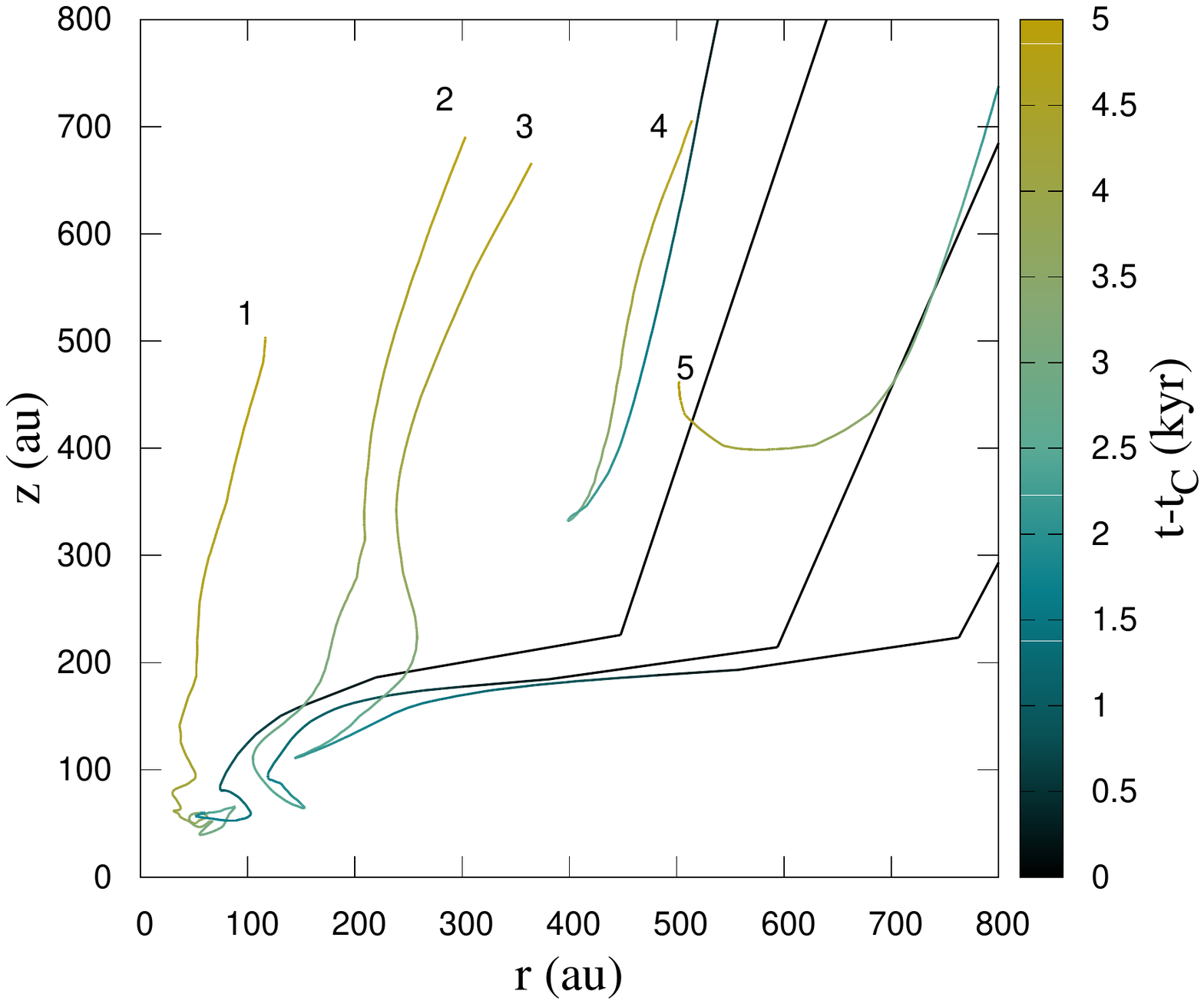}
  \caption{Specific AM of the gas in the outflow region in the plane $x=0$, for M5a3mrn. The thicker black line represent the boundaries of the outflow at $t-t_\mathrm{C}=5$ kyr. Thin lines are the trajectories, projected onto the yz-plane, of virtual tracer particles located in the outflow in the last output, and whose path is integrated backward in time. The 50 black particles are randomly distributed in the outflow, and the 25 white particles are distributed only in the region of high specific AM. The brown triangles indicate the location of the particles in the last output. For clarity, the abscissa of white trajectories represent the cylindrical radius and not the actual y-coordinate. The red countour represents the disk. The bottom panel shows the r/z trajectory of 5 selected particles, labeled 1 to 5. The color represents the time.}
  \label{Figtrajectory}
\end{center}
\end{figure}

Figure \ref{Figtrajectory} shows a slice at the outflow scale of the specific AM, for M5a3mrn. Most of the specific AM of the outflow is located in the bottom wall of the cavity, with values larger than $10^{21}$ cm$^2$ s$^{-1}$. This area seems connected to the outer layer of the pseudo-disk, that contains a similar specific AM, higher than in the mid-plane at the same cylindrical radius.
We have also plotted the projected trajectories of 75 virtual tracer particles, in thin black and white lines. The 50 black particles are randomly generated in the outflow in the final time-step, and the 25 white particles are are generated only in the high specific AM region. In post-processing calculations, we have integrated the particles trajectory backwards using the local flow velocity. 
This figure shows that the outflowing gas exclusively comes from the envelope and the pseudo-disk. There is almost no direct matter transport from the disk or the mid-plane to the outflow. 
In the early phases of the outflow, several particles fall near the axis in what we call ``the rotor region". This region is a small cylinder of 50-70 au radius and $\lesssim 10$ au height, that is located 50 au above the disk (see the bottom panel). The size of the rotor is controlled by the centrifugal radius of the gas. Gas that eventually outflows never goes closer to the disk than the rotor. Particles remain up to 4 kyr in the rotor, rotating one to several times around the axis at $\approx 2$ km s$^{-1}$. There, the radial Lorentz force builds up gradually. The particles eventually move away from the axis and are able to ``climb" along a magnetic field line. This occurs when the radial acceleration of the Lorentz force becomes larger than $\approx 0.003$ cm s$^{-2}$. The specific angular momentum of the particles does not significantly vary when confined in the rotor, and is independent from the time spent in this region.
At later stages, the outflow is fed in both gas and AM by the pseudo-disk, from its sides. In particular, white trajectories in Figure \ref{Figtrajectory} show that the high specific AM region of the outflow is supplied by the outer layers of the pseudo-disk, in which the specific AM is larger than in the mid-plane.
A fraction of the collapsing envelope is swept-up by the outflowing gas, which increases the outflow's mass. However, the swept-up gas holds a low AM, thus contributing negatively to the outflow's AM. Additionally, the gas at higher altitudes is launched at an earlier time, coming from more inner regions of the initial cloud, where the specific AM is lower. This explains why the specific AM decreases with height.
The bottom panel of Figure \ref{Figtrajectory} represents the radial-vertical path of 5 selected particles, labeled 1 to 5. Particle 1 arrives in the outflow within 1 kyr after the first Larson core formation, and stays in the rotor region few kyr before being ejected vertically and drifting radially as it ascends. Particles 2 and 3 enter the outflow at 2 kyr, but too late to be part of the rotor. Particles 4 and 5 arrive at a much later stage, after 3 kyr, and are immediately carried upward by the outflow.

\subsection{Ion-neutral drift}\label{Seciondrift}

The ambipolar diffusion is the drift between ions and neutral particles, at the origin of the decoupling between the gas and the magnetic field. The drift velocity is expressed by
\begin{equation}
  \mathbf{u}_\mathrm{AD} = \frac{c^2}{4\pi}\eta_\mathrm{AD} \frac{\mathbf{J}\times \mathbf{B}}{B^2},
\end{equation}
and the ion velocity is
\begin{equation}
  \mathbf{u}_\mathrm{ions} = \mathbf{u} + \mathbf{u}_\mathrm{AD},
\end{equation}
which is also the speed of the magnetic field lines motion because they are attached to the ions in the absence of the Hall effect.

\begin{figure}
\begin{center}
\includegraphics[trim=3cm 3cm 0cm 2cm,width=0.49\textwidth]{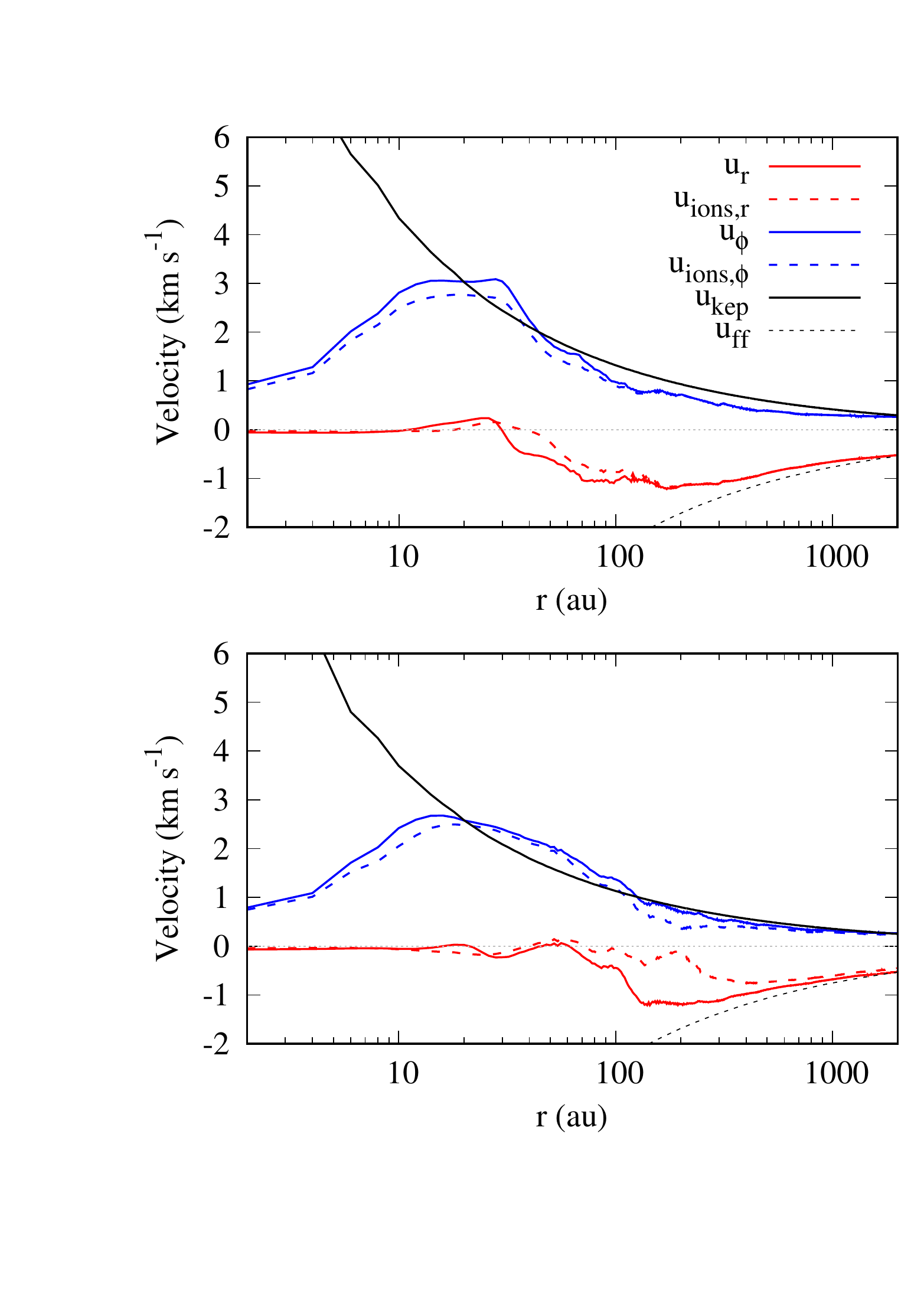}
  \caption{Azimuthaly averaged velocity profiles in the mid-plane for M5a3mrn (top panel) and M5a3big (bottom panel) at $t-t_\mathrm{c}=4$ kyr. Red curves represent the radial velocities, blue curves are the azimuthal velocities and the black curve indicate the Keplerian velocity. Solid lines represent the gas while dashed lines represent the ions.}
  \label{Figprofileions}
\end{center}
\end{figure}

In the mid-plane of the pseudo-disk, due to the pinching of the field lines, $B_r$ rapidly changes with $z$. For this reason, and because $B_z$ is the main component of the magnetic field, the current is mostly toroidal, so ions are expected to drift in the radial direction (perpendicularly to the current and the magnetic field). The magnetic field lines resist the drag from the gas towards the center and tend to "relax" if the decoupling is strong enough.
Figure \ref{Figprofileions} shows the velocity profiles of the gas, the ions and the Keplerian velocity for M5a3mrn (top panel) and M5a3big (bottom panel) in the mid-plane at $t-t_\mathrm{C}=4$ kyr. The decoupling is clear in both cases. For M5a3mrn, the ions travel up to 0.5 km s$^{-1}$ faster or slower than the neutral gas, and their decoupling is the strongest before the accretion shock, between $r=30$ to $r=100$ au, inside the disk. For $r<1000$ au, the ions are rotating at a slower speed of $~0.1 - 0.3$ km s$^{-1}$ than the neutral gas. The twisting of the field lines is therefore weaker than what it would be in ideal MHD, reducing the magnetic braking. In M5a3big, between $r=60$ and $r=200$ au, ions resist the drag from the gas and almost reach a zero radial velocity. This behavior corresponds to the simulations of \citet{2018MNRAS.473.4868Z} with a similar grain size distribution. The decoupling in the azimuthal direction is not as strong overall, similar to the standard MRN case, but extended to $r=400$ au.

\begin{figure*}
\begin{center}
\includegraphics[trim=1cm 0cm 1cm 0cm, width=0.95\textwidth]{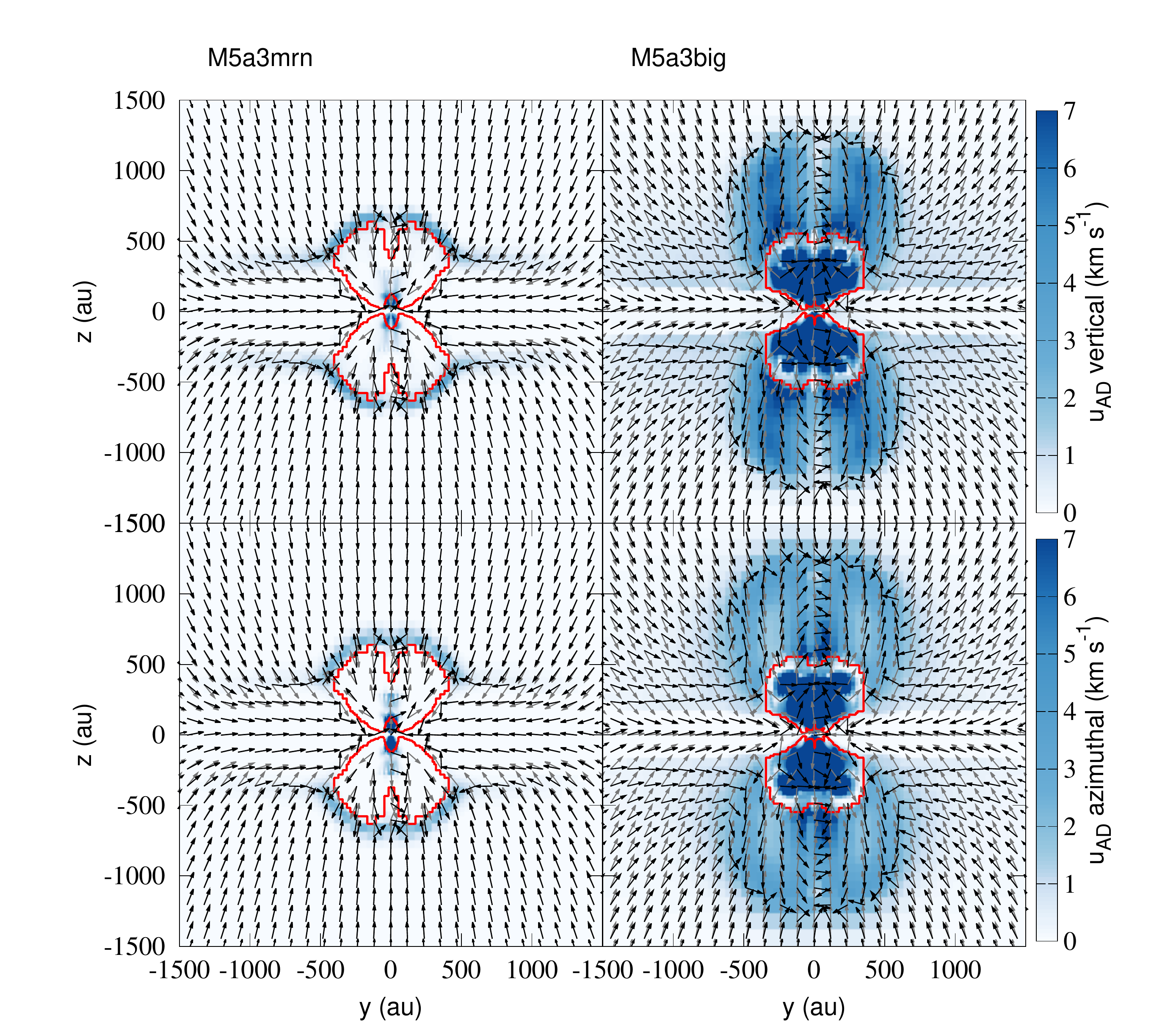}
  \caption{Relative velocities of the ions and the neutrals for M5a3mrn (left panels) and M5a3big (right panels) at $t-t_\mathrm{C}=2.5$ kyr. The color represents the drift velocity $||\mathbf{u}_\mathrm{AD}||=||\mathbf{u}-\mathbf{u}_\mathrm{ions}||$ along the z-direction (top panels) and $\phi$-direction (bottom panels). Grey arrows : neutral gas velocity direction, black arrows : ions velocity direction, red contour : neutral gas outflow.}
  \label{Figionoutflow}
\end{center}
\end{figure*}

We now look at the ion-neutral decoupling at larger scales, in the vicinity of the outflow. Figure \ref{Figionoutflow} displays the relative velocities of the ions and the neutrals for M5a3mrn and M5a3big at $t-t_\mathrm{C}=2.5$ kyr. The color represents the drift velocity $\mathbf{u}_\mathrm{AD}=\mathbf{u}-\mathbf{u}_\mathrm{ions}$ in the z and $\phi$ directions, while the white and black arrows indicate the direction of the neutrals and the ions respectively. Inside the pseudo-disk, the drift velocity is negligible, indicating a strong coupling, similarly to the velocity profiles in Figure \ref{Figprofileions} beyond few 100 au. In the infalling envelope however, the ions weakly decouple as they approach the mid-plane, until they enter the pseudo-disk. For $|z| \lesssim 500$ au, their trajectory is more bent toward the center than the neutrals. The decoupling is almost negligible in the envelope for the normal MRN distribution, but with the truncated distribution, the drift velocity reads amplitudes $\gtrsim 1$ km s$^{-1}$. In the outflow of M5a3big, the drift velocity reaches several $10$ km s$^{-1}$, similar to the value of the Alfven velocity, due to the extremely high ambipolar coefficient (two to four orders of magnitude higher than in the envelope or the disk).
However, the most striking features are the large zones of strong decoupling in the vicinity of the outflow for the truncated MRN case. Due to the high ambipolar diffusion, caused by the low density, and the strong electric current and magnetic field close to the central axis, the ions drift from the neutrals at velocities of several km s$^{-1}$. We even observe an outflow of ions outside and bigger than the actual previously-described outflow. Ions are escaping against the flow of infalling neutrals from the envelope at velocities of several km s$^{-1}$. The velocity difference between both kind of species exceeds 5 km s$^{-1}$ in these regions. The ions also strongly drift azimuthally at velocities of typically $2$ km s$^{-1}$. From now, to distinguish both structures, we call this decoupling region the ``ion-outflow", and the previously defined outflow as the ``neutral-outflow". To our knowledge, such behavior has never been reported in actual star forming clouds observations nor in simulations.
In the surrounding infalling envelope, ions also drift from the neutrals, though at lower speed, and seem to converge towards the ion-outflow in the same way that the neutral gas converges toward the neutral-outflow, as shown in Figure \ref{Figtrajectory}. 
This ion-outflow is actually a non-ideal magnetosonic wave driven by the neutral outflow, that is supersonic (Mach number $ \mathcal{M} = u/c_\mathrm{s} > 5$) but sub-Alfvenic ($\mathcal{M}_\mathrm{a}=u/c_\mathrm{A} < 0.5$). This wave propagates in the surrounding medium and accelerates ions ahead of the neutrals because of their much higher Alfv\'en speed. This phenomenon is countered by the ion-neutral drag, that is much higher in the standard MRN case, because a larger grain surface area leads to more collisions, which prevents the formation of a large precursor ion-outflow.
The drift velocity is correlated to the relative strengths of $B_\phi$ and $B_z$ shown in the bottom panels of figure \ref{Figangleb}. In M5a3mrn, $B_\phi/B_z$ is large ($> 2$) in the pseudo-disk and in the outflow, but weak ($<0.3$) in the envelope. The ratio is smaller in the pseudo-disk and the outflow for M5a3big, but it is larger in the envelope, with values close to $1$, where the drift velocity approaches $1$ km s$^{-1}$, and particularly important in the ion-outflow as $B_\phi/B_z>5$.

\section{Discussion} \label{SectDiscussion}

The question of the regulation of AM is deeply related to the formation of binary stars, particularly, in our non-turbulent case, by disk fragmentation. \citet{2008ApJ...677..327M} performed an extensive study on this subject by varying the magnetic field strength and the rotational velocity. They were able to define parameter ranges suitable for the formation of wide binary systems (separation of 3-300 au), close binary systems (0.007-0.3 au) and single stars. The value of our parameters falls within the wide binary region (see their figure 12), which is consistent with the high instability of the disks in our simulations (spiral arms $>$ 100 au, low Toomre's Q, fragmentation). The major differences are their initial condition, a Bonnor-Ebert sphere \citep{1955ZA.....37..217E,1956MNRAS.116..351B}, and the absence of ambipolar diffusion, with only the Ohmic diffusion. Nonetheless, in our study, a uniform sphere yields a higher accretion rate and the ambipolar diffusion weakens the magnetic braking, which promote the fragmentation and the formation of wide binaries. With a longer evolution, we could expect the formation of stable wide companions by disk fragmentation.
In their 3D simulations of protostellar collapse with ambipolar diffusion, \citet{2020A&A...635A..67H} found that the properties of the disk are weakly dependent on the initial conditions, except the magnetization but including the initial rotation velocity. This is a result that we concur. The disks are however heavily influenced by the accretion scheme on sink particles, which we do not have in this study. The presence of a sink particle reduces the mass and the extent of the disk, hence its ability to fragment. 

The relative importance between the magnetic braking and the outflow as an AM removal mechanism is still debated. The early model of \citet{2000ApJ...528L..41T} shows a dominance of the outflow over the magnetic braking in what would typically be the first Larson core phase. This work used the ideal MHD framework. Our study shows that ambipolar diffusion weakens the outflow even more than the magnetic braking. Therefore, the outflow may be able to dominate the magnetic braking in the absence of ambipolar diffusion. On the other hand, using 3D ideal MHD simulations, \citet{joos} finds that the magnetic braking transports more AM than the outflow. Depending on the orientation between the magnetic field and the rotation axis, the two effects are quantitatively closer than in our study, which confirms our hypothesis on the results of \citet{2000ApJ...528L..41T}. In our simulations, we find that at least 75\% of the AM accreted in the central region is removed by magnetic braking, even with high ambipolar coefficients. We can therefore estimate that the outflows (including later protostellar jets) cannot account for more than 25\% of the AM transport in the presence of ambipolar and Ohmic diffusion. At later stages, in an already formed disk without accretion, \citet{2019FrASS...6...54P} argue that the outflow should be the dominant process of angular momentum transport, because the Magneto-Rotational Instability (MRI) is greatly reduced by non-ideal MHD effects \citep{1994ApJ...421..163B,2013ApJ...769...76B}. Additionally, outflows driven by MRI can be significantly enhanced by the presence of the Hall effect \citep{Lesur2014}.

In Section \ref{Seciondrift} we report for the first time an ion-outflow precursor to the neutral gas outflow. With a typical ionisation of $10^{-7}$, the ion density in this region would be only of the order of $\approx 10$ cm$^{-3}$. However, with a velocity difference of several km s$^{-1}$ between ions and neutral, observations should be able to detect the precursor, using the CO molecule, a common tracer in protostellar outflows, and HCO$^+$, one of the most abundant ion and also a common tracer in young star systems. We also learn from Figure \ref{Figangleb} that a large ratio $B_\phi/B_z$ may be an indicator of a strong decoupling. \citet{2018A&A...615A..58Y} performed an observation of the ion-neutral drift in the Class 0 protostar B335 with the Atacama Large Millimeter Array (ALMA) telescope. They did not detect any strong decoupling down to the 100 au scale, finding an upper limit of 0.3 km s$^{-1}$ on the drift velocity in the collapsing gas. The absence of a significant decoupling would indicate a weaker ambipolar diffusion than in our simulations, which is confirmed by the small size of the disk, less than 10 au. Since a Class 0 protostar is more advanced than the systems we present in this study, another possibility is that the ion-outflow is a short-lived phenomenon that has already disappeared. In any case, more observations of the ion-neutral drift are needed to bring constraints on the chemistry at stake, and most importantly the grain size-distribution, and as a proof of the importance of non-ideal MHD.

There are limitations to the monofluid approximation that could affect the ion-outflow. The inertia of charged particles is neglected, meaning that the velocity of ions always immediately reaches the theoretical velocity calculated from the drift velocity. Additionally, ions could be subject to a chemical drag: charge transfer reaction between the outflowing ions and the infalling neutrals would effectively reduce the average velocity of the ion fluid. It is difficult to quantify these effects, but it is possible that they reduce the drift velocity. More accurate results necessitate a bifluid framework and chemical calculations on-the-fly. In appendix \ref{Appcollrates}, we discuss how the computation of the drift velocity and the ambipolar resistivity is an ill-defined problem in the monofluid formalism.

Figures \ref{Figdiskmassrad_a3} and \ref{Figdiskmassrad_a4} show that the disks can experience large fluctuations in mass and radius. The instability of the disks and their large spiral arms may cause their fragmentation. Clumps of gas  are  ejected  and  no longer fullfill  the  criteria to  belong to  the  disk,  which  decreases the disk mass and radius.  Gas is also accreted onto the  central  region, whose  higher  thermal  pressure  excludes  it  from  the  disk. Finally,  as  the first  core  does  not  collapse  due  to  the absence of sink particle and the use of a stiff barotropic EOS, some  gas  may  be  expelled from the core into the disk. Note, however, that these fluctuations may appear  even  in  the  presence of  a  sink  particle  \citep[see][]{2020A&A...635A..67H}.

\section{Conclusions} \label{SectConclusion}

We have performed 3D simulations of protostellar collapses to study the regulation of AM through the analysis of the disk, the outflow and the magnetic braking. Our parameters are the mass and the thermal support of the initial sphere of gas, and the grain-size distribution, either the MRN distribution or a truncated MRN distribution from which small grains have been removed.
Our main results are the following.

\begin{itemize}
  \item The mass of the dense core has little impact on the properties of the disk and the outflow during the first Larson core phase.
  \item A higher thermal over gravitational energy ratio $\alpha$ reduces the accretion rate which yields smaller and lower-mass discs, and reduces the AM transported by the outflow in the MRN case, but not in the truncated MRN case.
  \item The influence of the mass of the dense core and $\alpha$ on the magnetic braking in the disk is extremely limited, supporting the idea that the first core and the disk ``forget" some of the initial conditions.
  \item Removing the small grains produces slightly larger disks containing more angular momentum due to the higher ambipolar diffusion resistivity, but the difference is not significant. The outflow is however considerably weaker, transporting 10 times less AM compared to the MRN distribution, but with a larger opening angle.
  \item Most of the gas and AM of the outflow comes from the outer layers of the pseudo-disk.
  \item By the end of the first core phase, the magnetic braking is responsible for up to 80 percent of AM loss in the disk, and 75 percent (for the truncated MRN) to 90 percent (for the MRN distribution) in the disk + pseudo-disk.
  \item While not negligible, the transport of AM by the outflow is largely below the magnetic braking, by a factor 5 (for the MRN distribution) to 100 (for the truncated MRN).
  \item During the first Larson core stage, only 10\% of the accreted angular momentum makes it to or remains in the disk, and almost the entirety of the $90\%$ remaining escapes by magnetic braking or the outflow.
  \item In the truncated MRN simulations, the high ambipolar diffusion leads to an ion-outflow that is a precursor of the neutral-outflow. Ions travel at several km s$^{-1}$ against the infalling envelope, at larger scale than the neutral outflow. Observational evidences of such structures would be informative regarding the chemistry at stake in protostellar collapses.
\end{itemize}

The conclusions show that non-ideal MHD plays a significant role in the regulation of AM during the protostellar collapse, and that the results heavily depend on the chemical model, especially the grain-size distribution.

\section*{acknowledgments}
  The authors thank Kazunari Iwasaki and Mordecai-Mark Mac-Low for their thoughts and advices that greatly helped our analysis. We thank the anonymous referee whose report helped us deepen and contextualize our analysis. We acknowledge financial support from an International Research Fellowship of the Japan Society for the Promotion of Science (JSPS), number P17802, from the "Programme National de Physique Stellaire" (PNPS) of CNRS/INSU, CEA and CNES, France, and from the American Museum of Natural History. Computations were performed on the computer Occigen (CINES) under the allocation DARI A0020407247 made by GENCI. K. Tomida was supported by the Japan Society for the Promotion of Science (JSPS) KAKENHI Grant Numbers 16H05998, 16K13786, 17KK0091, 18H05440. K.Tanaka acknowledges support from NAOJ ALMA Scientific Research Grant 2017-05A, JSPS KAKENHI Grant JP19H05080, JP19K14760.

\appendix

\section{Ion-neutral collision rates}\label{Appcollrates}

In this section we analyse the influence of the drift velocity onto the ambipolar resistivity, because we deem important to emphasize some points regarding the calculation of the non-ideal MHD resistivities in the one-fluid approximation.
The detail of the formulae can be found in several studies \citep{1986MNRAS.218..663N,2016A&A...592A..18M}. Collision rates between ions and neutral are needed in the calculation, and most recent studies \citep{2016A&A...592A..18M,2016PASA...33...41W,2016MNRAS.460.2050Z} take these numbers from \citet{PintoGalli2}. The fitting formulae of their Table 2 are commonly used, and depend on the ion velocity
\begin{equation}
  v_\mathrm{rms}=\sqrt{v_\mathrm{d}^2+\frac{8kT_{ss'}}{\pi \mu_{ss'}}},
\end{equation}
where $v_\mathrm{d}$ is the drift velocity between species $s$ and $s'$, $k$ the Boltzman constant, $T_{ss'}$ the weighted average temperature, that we consider to be the fluid temperature in the one-fluid approximation, and $\mu_{ss'}$ the reduced mass of both species. In multi-fluid calculations, it is possible to derive $v_\mathrm{d}$ from the various species velocities. Since the quantity is not directly accessible in the one-fluid framework, it is often assumed to be zero. However, in cases of strong ambipolar drift, the drift velocity may very well exceed the thermal velocity, the second term. In this case, the approximations on collision rates, and resistivities, would prove inaccurate. Moreover, as detailed in their table 2, the fitting formulae of momentum transfer coefficients between ions and an H$_2$ are not valid if $v_\mathrm{rms}$ exceeds $5$ to $10$ km s$^{-1}$. We therefore have to be extremely careful in our analysis. In this paper, the drift velocity we compute in our simulations may exceed $10$ km s$^{-1}$, but always in the neutral-outflow. We therefore consider our results about the ion-neutral drift inaccurate om this region.

\begin{figure}
\begin{center}
\includegraphics[trim=3cm 2cm 3cm 1.5cm,width=0.49\textwidth]{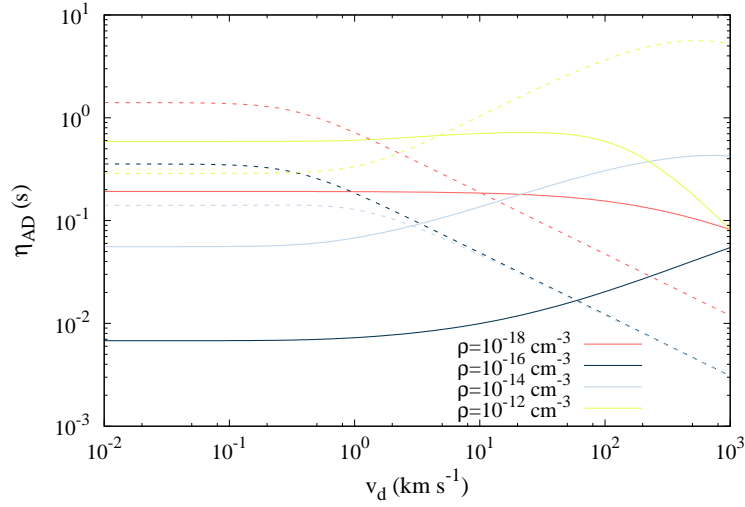}
  \caption{Ambipolar resistivity as a function of $v_\mathrm{d}$ for various densities. Red color: $\rho=10^{-18}$ g cm$^{-3}$, blue color: $\rho=10^{-16}$ g cm$^{-3}$, purple color: $\rho=10^{-14}$ g cm$^{-3}$, green color: $\rho=10^{-12}$ g cm$^{-3}$. The solid lines represent the MRN case and the dashed lines represent the truncated MRN case.}
  \label{Figetavdrift}
\end{center}
\end{figure}

We use the drift velocity $v_\mathrm{d}$ as a parameter and look at its influence on the ambipolar resistivity. We assume the same drift velocity for all molecular ions. Small grains ($< 1$ $\mu$m) being well coupled to the gas \citep[see for instance][]{2014MNRAS.440.2136L,2019A&A...626A..96L}, we do not expect them to significantly drift from H$_2$. Electrons have a high thermal velocity, their $v_\mathrm{rms}$ is thus not strongly affected by a drift velocity of a few km s$^{-1}$. Here, we have assumed $T=10$ K, and $B=1.43 \times 10^{-7} \sqrt{\rho/(m_\mathrm{p}\mu_\mathrm{p})}$ \citep{LiKrasnopolskyShang}.
Ambipolar resistivities as a function of the drift velocity are displayed in Figure \ref{Figetavdrift}, for various densities, and for both grain size distributions.

Below $v_\mathrm{d}=1$ km s$^{-1}$, the ambipolar coefficient does not seriously diverges from its value at $v_\mathrm{d}=0$. However, the difference is significant at higher velocities for the truncated MRN, while the MRN case shows a more reduced variation. At 10 km s$^{-1}$, the resistivities are larger by a factor 2 to 5 for the truncated MRN, and extends beyond one order of magnitude at 1000 km s$^{-1}$ (although it is unlikely that such high speed would be reached in the context of star formation). This difference comes from the fact that grains are the dominant charged species in the MRN case, while it is the ions in the truncated MRN case, because of the lower number of grains in this density range. As a consequence, with a truncated MRN, the resistivities are more affected by variation of the ions drift velocities.
In the outflow, the density is of the order of $\rho=10^{-18}$ g cm$^{-3}$ and the magnetic field $B\approx 0.01$ G, which is higher than the value given by our prescription, but the results are qualitatively the same.

Collapsing the ions and neutrals momentum and energy equations removes one degree of freedom in the system, as the ambipolar resistivity represents itself the drift between both fluids. This issue represents a limit of the monofluid description of non-ideal MHD.

\bibliography{MaBiblio}{}
\bibliographystyle{aasjournal}



\end{document}